\documentclass{article}
\usepackage{graphicx}
\usepackage[usenames,dvipsnames]{xcolor}
\DeclareGraphicsExtensions{.jpg,.pdf,.jpg}
\usepackage{amssymb}

\title{Noninvasive imaging of 3D micro and nanostructures by
topological methods}

\author{A. Carpio\thanks{Departamento de Matematica Aplicada,
Universidad Complutense, Madrid, 28040, Spain},
T.G. Dimiduk\thanks{Department of Physics, Harvard University,
Cambridge, MA 02138, USA},
M.L. Rap\'un\thanks{Departamento de Matem\'{a}tica Aplicada a la Ingenier\'{\i}a Aeroespacial,
ETSI Aeron\'auticos, Universidad Polit\'ecnica de Madrid, Madrid 28040,
Spain},
V. Selgas\thanks{Departamento de Matem\'aticas, Escuela Polit\'ecnica
de Ingenier\'{\i}a, Universidad de Oviedo, Gij\'on 33203, Spain}
}

\date{to appear in SIAM J. Imaging Sciences, 9(3), 1324-1354, 2016}

\begin{document}
\maketitle

\begin{abstract}
We present  topological derivative and energy based procedures for the imaging of micro and  nanostructures using one beam of visible light of a single wavelength. Objects with diameters as small as $10$ nm can be located, and their position tracked with nanometer precision. Multiple objects distributed either on planes perpendicular to the incidence direction or along axial lines in the incidence direction are distinguishable.  More precisely, the shape and size of plane sections perpendicular to the incidence  direction can be clearly determined, even for asymmetric and non-convex scatterers.  Axial resolution improves as the size of the objects decreases.
Initial reconstructions may proceed by glueing together 2D horizontal slices between axial peaks or by locating objects at 3D peaks of topological energies, depending on the effective wavenumber.
Below a threshold size, topological derivative  based iterative schemes improve initial predictions of the location, size and shape of objects  by postprocessing fixed measured data. For larger sizes, tracking the peaks of topological energy fields that average information from additional incident light beams seems to be more effective.
\end{abstract}




\section{Introduction}
\label{sec:intro}

Ideally, imaging biological samples resolves structures in three dimensions, over a wide range of scales without harming the specimen. To prevent damage, live cells have traditionally been imaged using light microscopes. Visible light corresponds to a wavelength range of $380$-$740$ nanometers (nm).
Cells are large compared to the illumination wavelength, but comprise finer structures down to molecular sizes, see Table \ref{table1}.

Unstained biological samples lack contrast in straightforward optical imaging. To gain contrast on structures of interest, biologists have very productively used fluorescent dyes. However, such stains are invasive, phototoxic and tend to bleach over extended periods of time \cite{fluorotoxic}. For this reason, it is useful to consider techniques which can image biological structures in their native state. Living cells are almost transparent, affecting mostly the phase of light passing through them and making only minimal amplitude changes.  Differential interference contrast (DIC) and phase contrast microscopes are sensitive to these phase changes \cite{pc}, and have been the gold standard for optical imaging of unstained cells. More recent holographic techniques sense phase in a more quantitative manner that contains three dimensional information. While optical reconstructions reproduce an aspect of an object,  digital holography recreates object data from the recorded light measurements. 

Different techniques have been exploited to heighten spatial resolution of optical microscopes beyond
the diffraction limit \cite{born}. Confocal 3D imaging \cite{confocal} 
resorts to point illumination. The need to scan an imaging point
through the sample in all three dimensions limits its ability to capture fast 
events subjecting samples to substantial photon flux and potential photodamage. On the other hand, $4\pi$ laser scanning fluorescence microscopes \cite{4pi} use two objectives to divide the diffraction limit by 
a factor  $2$.  For visible light wavelengths, this limit ranges from about $68$ nm to $132$ nm. Improved resolution (down to $10$ nm) of cellular nanostructures marked with fluorescent stains has been achieved combining spectral precision distance microscopy and spatially illuminated modulation, that blends two or three different emitted wavelengths \cite{super2,super1}.

\begin{table}[htbp]
\centering \footnotesize
\begin{tabular}{|c|c|c|c|c|}
\hline
 Red blood cell & Nucleus & Chloroplast & Mitochondrion  \\
  \hline
 9  $\mu$m &  6 $\mu$m & 5 $\mu$m  &  3 $\mu$m \\
  \hline \hline
 Bacterium E. Coli & Membrane & Protein & Virus  \\
  \hline
 2  $\mu$m & 3-10  nm  & 10  nm & 5-300  nm \\
 \hline
\end{tabular}
\caption{Size  of cells and cellular elements.}
\label{table1}
\end{table}

Holographic techniques provide a noninvasive alternative
for high speed 3D live cell imaging \cite{holocell,holo}. 
Holograms can be recorded as fast as cameras permit, usually in the millisecond or microsecond range, and the process does not damage
samples.
An hologram is not an image, but an encoding of the light field  as an interference pattern, with information about both the intensity and the 
phase of the light field \cite{holoamplphase}.  Holograms can be 
decoded optically or numerically. 
Digital holography is devoted to the numerical reconstruction of digitally recorded  holograms  \cite{holodigital,holo}. Numerical processing of the encoded information allows us to compute simultaneously the phase and the intensity of the propagated wave field. It also makes it possible to focus on different object planes without resorting to optomechanical motion \cite{holosection,holosection2}, and postprocess the information
to improve resolution and remove aberrations \cite{holoimprovedepth,holoenhanced}. Resolution can be further 
enhanced by combining light incoming from different directions \cite{holomultipled}.
Inverse scattering analysis is employed in Ref. \cite{holotom} to track 
the dynamics of wavelength-scale particles with nanometer-scale 
precision and millisecond temporal resolution. 
These inverse scattering techniques are computationally intensive, 
but this burden can be reduced without loss of resolution by using a smaller number of randomly arranged detectors, as discussed in Ref. \cite{holotom2}.
Progress in digital holography is strongly related to advances in the mathematical techniques to decode light measurements.
In this paper, we demonstrate the potential of topological  methods  
for the noninvasive imaging of low contrast structures displaying a variety
of lengths in the nano and microscales using light measurements
in  holographic settings.

\begin{figure}
\centering
\includegraphics[width=8cm]{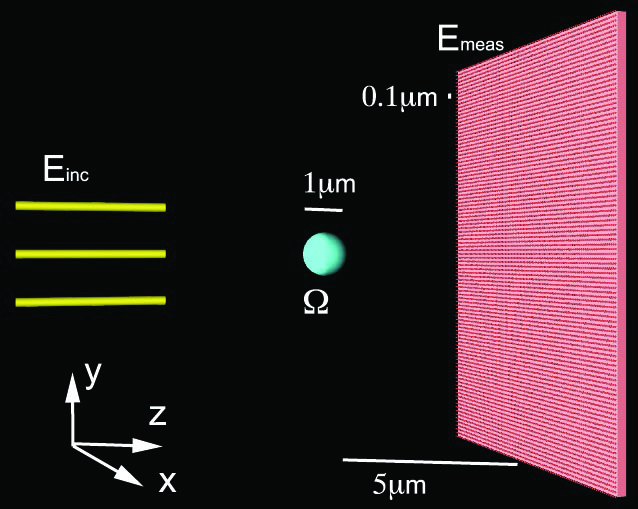}
\caption{Imaging setting. An object $\Omega$ is illuminated by a light beam $E_{inc}$. The total field $E$ is measured at a grid of detectors placed on a screen, orthogonal to the beam direction $z$. Typical sizes for the object range from $2\, \mu {\rm m}$ to $10$ nm. Its distance to the screen is about $5 \, \mu {\rm m}$. The screen size is $10 \, \mu {\rm m} \times 10 \, \mu {\rm m}$. The step for the grid of detectors is $0.1\, \mu {\rm m}$. Detectors correspond to a CMOS camera pixels
\cite{sheng}.}
\label{fig1}
\end{figure}

Topological methods recast the inverse light scattering problem as a 
constrained optimization problem and analyze the potential of the 
topological energy and the topological derivative of the resulting shape 
functional to image the objects.
To simplify, we assume that measurements of the full light wave field  
$E$  are available at set of detectors, see Fig. \ref{fig1}. 
The magnitude actually measured in holograms is 
$|H|^2 = |E+R|^2 = |E|^2 + E \overline{R} + \overline{E} R + |R|^2$, 
where $R$ is an independent reference light beam, assumed to be 
a uniform plane wave, whereas $E$ is the wave field generated when 
the incident light $E_{inc}$ interacts with the object.
Notice that the quantity we are interested in is $E$, the complex 
(magnitude and phase) electric field. When imaging biological or 
similar samples, reasonable approximations might allow one to extract 
the electric field from the recorded hologram.
For example,  for inline holography, we first normalize by 
the constant reference field:
\begin{equation}
  \left|\frac{H}{R}\right|^2 = 1 + \frac{E}{R} + \frac{\overline{E}}{\overline{R}} + \left|\frac{E}{R}\right|^2.
\end{equation}
Since cells and many other objects we are interested in imaging are weak scatterers, it is a reasonable approximation to neglect the $|E/R|^2$ term. Thus we can write:
\begin{equation}
 \left|\frac{H}{R}\right|^2 -1 \approx \frac{E}{R} + \frac{\overline{E}}{\overline{R}}.
 \label{aproximation}
\end{equation}
Effective algorithms  to extract the total field $E$ from approximation (\ref{aproximation}) are yet to be devised. We assume here that
$E$ is known at the detectors.

Topological derivative based methods have been implemented in many 
inverse scattering problems,  see \cite{carpiorapun1,carpiorapun3,caubetfluid,feijoo,guzinaacoustic, guzinaelastic} for applications in acoustics, elasticity, fluids and tomography, for instance, usually in  two-dimensional settings.
They have the ability of providing first guesses of objects in absence of any a priori information. Such approximations can then be improved by iterative
schemes, as we will discuss later.
The performance of topological derivatives usually varies with the wavenumber, as observed in acoustic scattering problems. For small wavenumbers, regions where topological derivatives take large negative values mark object locations. As the wavenumber increases, the topological derivative fields develop oscillations.  Remarkably, for large wavenumbers and two-dimensional problems, large negative values mark the boundary of single objects when a wide distribution of incident waves and receptors is considered \cite{carpiorapunln}. Moreover, for large wavenumbers and in three dimensions, topological derivative fields that average information provided  by incident waves from all possible directions allow us to reconstruct the boundary of  isolated convex objects \cite{guzinalarge}. However, handling multiple or  non-convex objects becomes a challenge. 
To tackle difficulties caused by oscillatory behavior,  topological derivatives have been replaced by topological energies  in two-dimensional  time-dependent wave scattering \cite{topologicalenergy}.

In this paper we mostly  consider a single incident direction because that corresponds to the way optical microscopes are conventionally built.
It would be possible for a custom built microscope to use some range of 
incident directions, but that be fairly sharply limited by the numerical 
aperture of the observing objective lens. 
Here, we consider time-harmonic light scattering in three dimensions (3D) 
in the imaging setting represented in Figure \ref{fig1}.
The wavenumbers in our specific imaging setting,  detailed in Sections \ref{sec:setting} and \ref{sec:3D1st}, lie in the {\em transition to large wavenumbers}.
To overcome difficulties created by the  aforementioned oscillatory behavior of the topological derivative, we analyze information from both topological energies and topological derivatives.
We are able to locate {\em multiple  three-dimensional} objects and spot the shape of {\em non-convex} and {\em asymmetric} objects,  down to spatial scales of {\em a few nanometers} using just {\em one incident wave} and {\em one detector screen.} {\em Objects in the shadow} of other objects in the light direction are clearly distinguished.  As the size of the objects varies, a transition in the behavior of the topological fields is observed.  For sizes below a micron, objects are located at  peaks of these 3D fields. For larger sizes this remains true only in lateral 2D slices. On axial 2D slices, the topological fields concentrate at the edges, before and past the object.  Interestingly, some holographic reconstruction procedures show that the brightest points become focal 
caustics of objects as they get larger. Essentially the object can behave as a lens, and such reconstructions pick out the bright point where the object 
brings the light into smaller focus \cite{lensing1,lensing2}.
A similar effect might be occurring with topological fields here.  
Ref. \cite{guzinalarge} contains a further theoretical discussion
assuming impenetrable defects and short wavelengths.

Proposing a deterministic (no or minimal human input needed) algorithm for turning topological fields into 3D models of objects would be of great value.
For small enough objects, we are able to improve a first guess of the scatterers employing an iterative procedure that postprocess the same measured data.
On the contrary, our preliminary tests suggest that objects larger than the wavelength 
are better resolved by considering additional incident waves. 
From the computational point of view, implementing iterative schemes involving shapes requires overcoming a number of  nontrivial technical issues about how to generate and mesh the   topological field based three-dimensional approximations  of the obstacle.  Such  approximations must be designed to adjust objects defined by the knowledge of which points of a  three-dimensional cubic  grid  lie inside and outside. We propose a way to deal with this problem which is well suited for  some coupled finite element-boundary element  solvers.
Future work could investigate how much can be gained by varying incident waves within the acceptance angle of the objective accounting for the asymmetric loss of high frequency information along the displaced angle. 

 Topological field based methods provide first guesses of objects in absence of a priori information, other than the measured data, the incident waves, and the material parameters of the background medium.  They locate abrupt changes in the nature of the medium. 
Once an initial guess of the contours of the scatterers is available, we have two choices. First, we may assume that their material parameters are known and seek to improve the knowledge of their geometry. This can be done by topological field based iterative schemes \cite{carpiorapun1}, as discussed
above, or employing
other reconstruction methods to track sharp interfaces in the parameter values (identified with contours of objects), such as level sets \cite{oliver1,oliver2} or contour deformations \cite{caubetfluid}. Both are often initialized using topological derivatives.
If the material parameters of the anomalies are unknown,  we may combine these techniques with gradient methods to approximate the spatial variation 
of such parameters within the objects \cite{carpiorapun2,carpiorapun3}. Alternatively, we 
may apply from the beginning  strategies to reconstruct the parameter profile distribution everywhere. 2D studies in an electric impedance tomography 
setting \cite{carpiorapun3} suggest that first defining object boundaries with topological derivative methods and then approximating spatial variations of the unknown object parameters with a gradient technique is more efficient than employing the simple gradient technique to reproduce spatial variations everywhere from the beginning. In microwave imaging with sources distributed all around the objects, sophisticated iterative minimization schemes  yield
good reconstructions of permittivity profile distributions in a number of
specific tests \cite{capatano,chaumet,eyraud,li,yu,zaeytijd} for objects  smaller than the wavelength.  They are all Newton type algorithms. Linear  sampling 
has also been exploited to reproduce the morphology of these targets  
surrounded by sources \cite{capatano}.
In our setting, we are restricted to one source, as mentioned above, 
and the refractivity indexes (therefore the objects permittivities) are often known. 
However, it would be of interest to investigate the performance
of these Newton schemes combined with topological methods for an
analysis of  the permittivity profiles within the objects, as it might help to 
resolve structures in more detail. 

In our imaging setting, considering multiple frequencies in the visible light 
range does not seem to produce significative improvements over working 
with one, as deduced from the numerical simulations we have performed 
so far. On the contrary, averaging information from multiple frequencies has been shown to enhance precision in  3D microwave imaging \cite{yu} and  2D  acoustic applications \cite{multifrequency,park}.
Let us emphasize that, in holography, light detectors are placed past the objects, whereas in underground acoustic imaging, for instance, emitters 
and receptors are usually placed on the same surface.  In
microwave imaging,  sources  are often distributed all around the object while receivers are restricted to a circle around it. These variations strongly affect the oscillatory structure of the resulting three-dimensional topological derivative and topological energy fields. 

The paper is organized as follows. Section \ref{sec:setting} describes the imaging setting  and formulates the associated constrained optimization problem. Section \ref{sec:topological}
introduces topological derivatives and energies as effective tools in the imaging of cellular structures using visible light. Section \ref{sec:3D1st} illustrates the ability of these techniques to track the size, shape and location of three-dimensional rods and balls of varying size, as well as multiple, non-convex and asymmetric objects.  Section \ref{sec:implementation} develops iterative strategies to improve the initial reconstructions. We discuss the procedures used to construct  three-dimensional approximations of the obstacles and to solve the auxiliary boundary value problems for wave fields.
Section \ref{sec:shape} illustrates how predicted shapes may improve depending on their size, either by iteration or by adding incident waves. Finally, Section \ref{sec:discussion} summarizes our conclusions.

\section{Inverse scattering problem}
\label{sec:setting}

We plan to image objects of cell size using light.
Cells and cell structures have usually sizes in the micrometer range 
($1\, \mu$m= $10^{-6}$ m= $10^3$ nm), see Table \ref{table1}.
As mentioned earlier, the ability of an imaging system to resolve detail is ultimately limited by diffraction (interference of waves when a wave encounters an obstacle comparable in size to its wavelength).
The resolution of light microscopes is limited by the wavelength of visible light, which is in the range $400-700$ nm. According to Table \ref{table1}, this should make possible to visualize cells, and some structures within the cell, such as the nucleus, mitochondria and chloroplasts.  On the contrary, to resolve for viruses or finer cell details enhanced resolution is needed. Similarly, the cell membrane is only about $3-10$ nm and cannot be resolved with standard light microscopes.  Bacteria usually do not have any membrane-wrapped organelles (e.g., nucleus, mitochondria, endoplasmic reticulum), but they do have an outer membrane.

\begin{figure}
\centering
\hskip 1cm (a) \hskip 5cm (b) \\
\includegraphics[height=5cm]{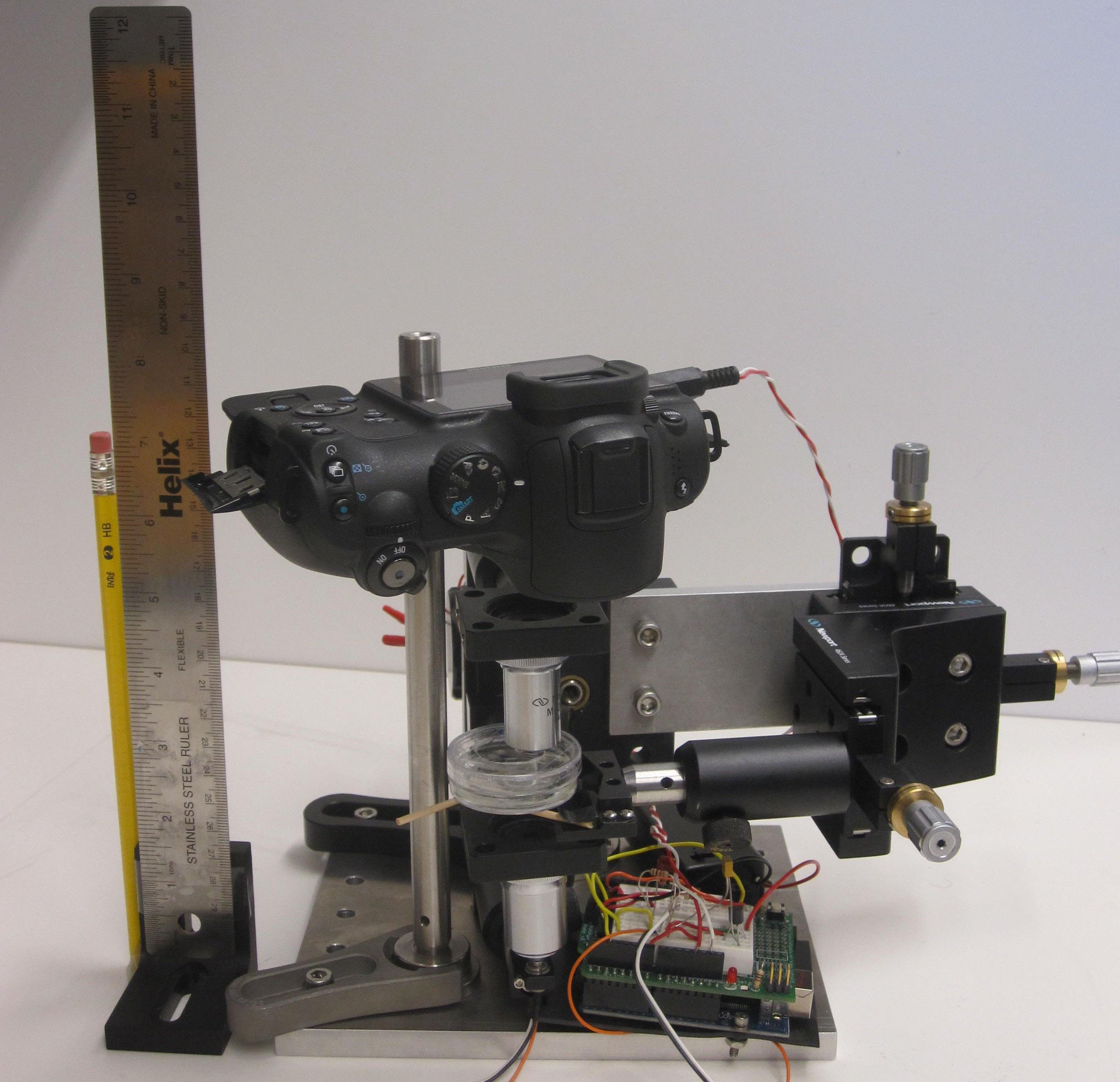}
\hskip 1cm
\includegraphics[height=5cm]{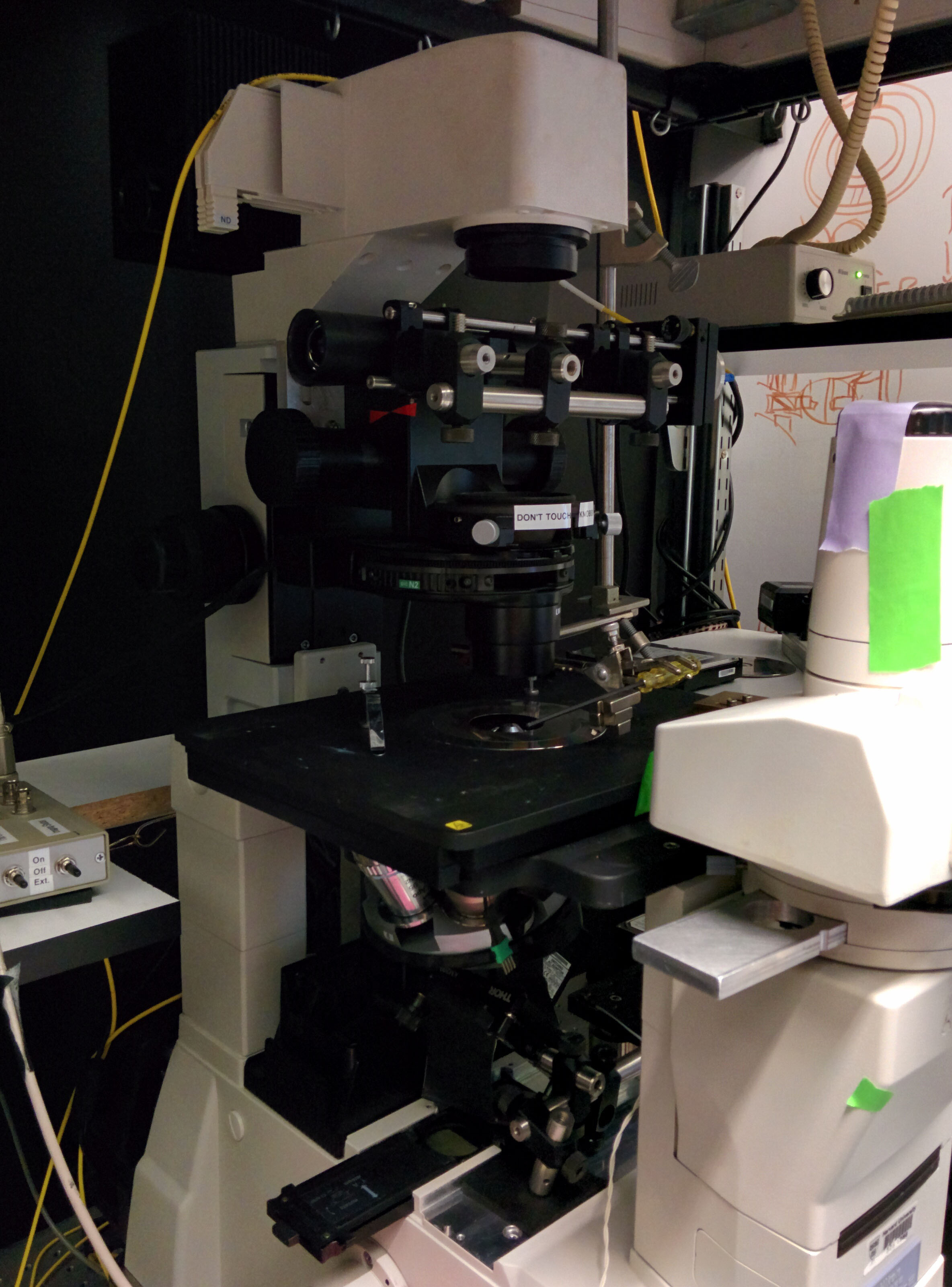}
\caption{Holographic microscopes used for biological and soft matter imaging at Harvard University. (a) Small, self-contained, battery powered microscope for portable and in incubator use (b) Full-size, high precision microscope with optical trapping and conventional imaging capabilities.}
\label{fig1bis}
\end{figure}

The imaging setting is depicted in Figure \ref{fig1}.  Holographic microscopes based on a similar setting are shown in Figure \ref{fig1bis}.
In such settings, cells are placed near a screen illuminated by an incident wave  ${\cal E}_{inc}$. Part of the incident wave is transmitted inside the objects $\Omega$, and the rest is scattered. The evolution of the light wave field
\[
{\cal E}(\mathbf{x},t):=\left\{\begin{array}{ll}
{\cal E}_{inc}(\mathbf{x},t)+{\cal E}_{sc}(\mathbf{x},t),&\quad\mbox{$\mathbf{x}\in\mathbb R^3\setminus\overline{\Omega}$},\ \ t>0,\\
{\cal E}_{tr}(\mathbf{x},t),&\quad\mbox{$\mathbf{x}\in\Omega$}, \ \ t>0,
\end{array}
\right.
\]
is governed by a wave equation
\begin{equation}
\varepsilon(\mathbf{x}) {\cal E}_{tt}(\mathbf{x},t) -
{\rm div} ({1\over {\mu(\mathbf{x})}} \nabla {\cal E}(\mathbf{x},t)) = 0,
\label{ec:vble}
\end{equation}
where $\varepsilon$ is the permittivity and $\mu$ is the permeability. 
 Electromagnetic wave equations are used to fit
real holograms in references \cite{fung,kaz,lee,perry,holotom}. 
The electric field scattered by spherical
particles is  computed exactly using  the Mie solution in \cite{kaz,lee}, 
and extensions of the Mie solution to multiple particles in
\cite{fung,perry}. Reference \cite{holotom} exploits the DDA (discrete dipole 
approximation)  to solve the wave equation for more general shapes
and resorts to less sophisticated inversion techniques to work with real 
holograms of simple objects.

If we assume that $\varepsilon$ and $\mu$ are constant inside the exterior 
medium and inside the objects $\Omega$,
\[
\varepsilon(\mathbf{x})=\left\{\begin{array}{ll}
\varepsilon_e,&\quad \mathbf{x}\in\mathbb R^3\setminus\overline{\Omega},\\
\varepsilon_i,&\quad \mathbf{x}\in\Omega,
\end{array}
\right.\qquad
\mu(\mathbf{x})=\left\{\begin{array}{ll}
\mu_e,&\quad \mathbf{x}\in\mathbb R^3\setminus\overline{\Omega},\\
\mu_i,&\quad \mathbf{x}\in\Omega,
\end{array}
\right.
\]
we have wave equations of the form
\begin{equation}
 {\cal E}_{tt} - c^2 \Delta {\cal E} = 0,
\label{ec:constant}
\end{equation}
with constant wave speeds $c_e={1\over \sqrt{\varepsilon_e \mu_e}}$ in the exterior medium $\mathbb R^3\setminus\overline{\Omega}$ and $c_i={1\over \sqrt{\varepsilon_i \mu_i}}$ inside $\Omega$.
If we further assume that the incident wave is time-harmonic,
\[
{\cal E}_{inc}(\mathbf{x},t)=\mbox{Re}[e^{-2\pi\imath \nu t} E_{inc}(\mathbf{x})],
\]
where $\nu$ is the frequency of the light wave (in  Hz or cycles/second), then, the wave field $\cal E$ is time-harmonic as well,  ${\cal E}(\mathbf{x},t)=\mbox{Re}[e^{-2\pi\imath\nu t} E(\mathbf{x})]$, and we arrive at the following stationary problem for the (complex) amplitude $E(\mathbf{x})$:
\begin{equation}\label{HelmholtzDim}
\left\{\begin{array}{ll}
\Delta E+\big(\frac{2\pi\nu}{c_e}\big)^2E=0,&\quad\mbox{in $\mathbb R^3\setminus\overline\Omega$},\\
\Delta E+\big(\frac{2\pi\nu}{c_i}\big)^2E=0,&\quad\mbox{in $\Omega$},\\
E^-=E^+,&\quad \mbox{on $\partial\Omega$},\\
\frac1{\mu_i}\, \partial_{\mathbf{n}} E^-=\frac1{\mu_e}\, \partial_{\mathbf{n}} E^+, &\quad \mbox{on $\partial\Omega$}.
\end{array}
\right.
\end{equation}
The superscripts ``$+$" and ``$-$" at the transmission conditions on the boundary $\partial \Omega$ denote limits from the interior and the exterior of $\Omega$ respectively, and $\partial_\mathbf{n}$ stands for normal derivatives, with $\mathbf{n}$ pointing outside $\Omega$. The problem is completed with the standard Sommerfeld radiation condition at infinity,
\begin{equation}\label{Sommerfeld}
\lim_{r:=|\mathbf{x}|\to \infty} r\big(\partial_r (E-E_{inc})-\imath\frac{2\pi\nu}{c_e}(E-E_{inc})\big)=0,
\end{equation}
modeling that only outgoing waves are allowed, that is, waves are not scattered from infinity.

For simplicity, wave amplitudes and distances are nondimensionalized in terms of a reference length $L$. Since we plan to use incident waves of different frequencies (therefore associated to different wavelengths), we choose a reference length of the order  of magnitude of the size of the objects we wish to locate: $L= 1\,\mu$m.  Setting $\mathbf{x}':=\mathbf{x}/L$,  $E':=E/a$ (the factor $a$ represents the units of the electric field) and using the notation \[\Omega':=\Omega/L, \qquad E'(\mathbf{x}'):=E(\mathbf{x})/a,\qquad E'_{inc}(\mathbf{x}'):=E_{inc}(\mathbf{x})/a,\] and the dimensionless parameters
\begin{eqnarray}
\beta:=\frac{\mu_e}{\mu_i},\qquad k_e:=\frac{2\pi\nu L}{c_e},\qquad k_i:=\frac{2\pi\nu L}{c_i},
\label{kL}
\end{eqnarray}
problem (\ref{HelmholtzDim})--(\ref{Sommerfeld}) is written as:
\begin{equation}\label{forward}
\left\{\begin{array}{ll}
\Delta E'+k_e^2 E'=0,&\quad\mbox{in $\mathbb R^3\setminus\overline{\Omega'}$},\\
\Delta E'+k_i^2 E'=0,&\quad\mbox{in $\Omega'$},\\
(E')^-= (E')^+,&\quad \mbox{on $\partial\Omega'$},\\
\beta (\partial_{\mathbf{n}}  E')^-= (\partial_{\mathbf{n}} E')^+, &\quad \mbox{on $\partial\Omega'$},\\
\displaystyle\lim_{r:=|\mathbf{x}'|\to \infty} r\big(\partial_r (E'- E'_{inc})-\imath k_e(E'- E'_{inc})\big)=0.
\end{array}
\right.
\end{equation}
The magnetic permeability of  tissues is almost the same as in the free space 
$\mu_0$, and does not change with  frequency \cite{lin}; therefore, we set $\beta=1$.

We work with laser lights of wavelengths $405 $ nm (violet) and $660 $ nm (red), in the lower and upper limits of the visible wavelength range.
The corresponding frequencies are   $7.38e14 $ Hz  for  light of wavelength $405 $ nm  and $4.5e14$ for  $660 $ nm light. The selected wave speeds are $c_e=2.25e8 \, {\rm {m\over s}}$  in the background medium and $c_i=2.15e8 \, {\rm {m\over s}}$ or $c_i=1.87e8 \, 
{\rm {m\over s}}$ inside the  objects.
These choices yield the dimensionless reference wavenumbers collected in Table \ref{table3} for a reference length $L= 1\,\mu {\rm m}$. We will fix these values in the numerical simulations for different object sizes ${\cal L}$. However, each simulation has  `effective' dimensionless wavenumbers 
${\cal K}_e= k_e {\cal L}/L$  and ${\cal K}_i= k_i {\cal L}/L$. 

\begin{table}[hbtp]
\centering \footnotesize
\begin{tabular}{|c|c|c|c|c|}
\hline
  & Violet light ($405$ nm) & Red light ($660$ nm)  \\
  \hline
$k_e$ (background medium) &  $20.60$  & $12.56$  \\
  \hline
$k_i$ (lipid membranes) & $24.79$ & $15.12$  \\
  \hline
$k_i$ (organelles, nucleus) & $21.56$  & $13.15$  \\
 \hline
 $\beta$  &  1 & 1 \\
 \hline
\end{tabular}
\caption{Dimensionless parameters for the model.}
\label{table3}
\end{table}

To image the object $\Omega$, we locate a mesh of detectors 
$\{\mathbf x_1,\dots,\mathbf x_N\}$ on a plane screen $\Pi_{meas}$ 
that is illuminated 
by a plane time-harmonic incident wave in the direction $\mathbf d$ orthogonal to the screen. To simplify, we assume that $\mathbf d=(0,0,1)$ (see Figure 1). Mathematically, this is modeled by
\[E_{inc}(\mathbf{x})/a=E'_{inc}(\mathbf{x}')=E'_{inc}(x',y',z')=
e^{\imath k_e z'}.\]
The total field $E$ is then measured at the detectors, obtaining the set of data
$E_{meas,j}$ for $j=1,\dots,N$. The measured data encode information about the position, structure, size, and shape of $\Omega$. Extracting this information from the data constitutes the inverse problem to be solved: find the  object $\Omega$ such that
\[
E(\mathbf{x}_j)=E_{meas,j},\qquad j=1,\dots,N,
\]
where  $E$ is the solution of the transmission problem (\ref{HelmholtzDim})--(\ref{Sommerfeld}). Equivalently, in terms of the dimensionless formulation, the
problem reads as: find the domain $\Omega'$ such that
\begin{equation}\label{ip}
E'(\mathbf{x}'_j)=E'_{meas,j},\qquad j=1,\dots,N,
\end{equation}
where  $E'$ is the solution of the transmission problem (\ref{forward}).

This inverse problem is nonlinear with respect to $\Omega$ and strongly ill-posed \cite{Colton}. Uniqueness is  theoretically proved when information from all space directions is available \cite{alldirections}.
However, we are limited to consider just one incident direction.
Furthermore, even if uniqueness were guaranteed, the problem would still be 
ill-posed because  $\Omega$ does not depend continuously on the data, i.e., slight changes in the data may produce large variations in $\Omega$.  In consequence, instead of imposing the identities (\ref{ip}), which is highly demanding, we reformulate the inverse problem in a weaker form, recasting 
the inverse scattering formulation as an optimization problem. This allows us 
to devise computational strategies to reconstruct the objects. In addition, it provides a more regular formulation to deal with the ill-posedness of the problem.  Thus, now we seek  objects $\Omega'$ minimizing:
\begin{eqnarray}
J({\mathbb R}^3 \setminus \overline{\Omega'})=
\frac12 \sum_{j=1}^N
|E'(\mathbf{x}'_j)-E'_{meas,j}|^2, \label{cost}
\end{eqnarray}
where $E'$ is the solution of the transmission problem (\ref{forward}) with objects $\Omega'$ and $E'_{meas,j}=E_{meas,j}/a$.
The  boundary value problem for the stationary wave equation (\ref{forward}) acts now as a constraint in the minimization problem. Notice that the cost functional (\ref{cost}) vanishes for the true scatterers, which furnish a global minimum. We will use this dimensionless formulation hereinafter. The superscripts $'$ will be omitted in the sequel for ease of notation.

\section{Topological field based imaging}
\label{sec:topological}

Strategies to solve the optimization problem described in the previous section vary depending on the a priori knowledge.  For example, information about the number or shape of the objects may simplify the procedure. We will address here the general problem with no a priori information, showing that topological energies and derivatives are effective tools for light imaging of cellular samples.

Topological derivatives of cost functionals with a mathematical structure analogous to functional (\ref{cost}) have been used to predict the structure of scatterers, though in different imaging settings, see \cite{carpiorapun1,feijoo,guzinaacoustic} for instance.
The topological derivative of a shape functional $J({\cal R})$, defined in a region ${\cal R}\subset \mathbb R^3$, measures its sensitivity when an infinitesimal object is located at each point $\mathbf x\in \cal R$  \cite{Sokowloski}. In case the infinitesimal object is a ball of radius $\varepsilon$ centered about $\mathbf x$, we have the expansion
\begin{equation}\label{dteout}
J({\cal R} \setminus \overline{B_\varepsilon (\mathbf{x})})=J({\cal
R})+{4\over 3} \pi \varepsilon^3 D_T(\mathbf{x},{\cal R})+o(\varepsilon^3),\qquad
\varepsilon\to 0.
\end{equation}
$D_T(\mathbf{x},{\cal R})$ is the topological derivative (or sensitivity) of the cost functional at ${\mathbf x}$. Whenever $D_T(\mathbf{x},{\cal R})<0$,  we deduce $J({\cal R}\setminus \overline{B_\varepsilon (\mathbf{x})})<J({\cal R})$ for $\varepsilon>0$ small. If we place objects in regions where the topological derivative takes large negative values, the error functional is expected to decrease, yielding a prediction of the number, size and location of the scatterers.

An explicit expression for the topological derivative of functional (\ref{cost}) in terms of adjoint and forward fields is given in \cite{carpiorapun1,guzinaacoustic}. For any ${\mathbf x} \in{\cal R}$
\begin{eqnarray}
D_T(\mathbf{x},{\cal R}) =
{\rm Re} \left[ 3{1 - \beta \over 2 + \beta}
\nabla E(\mathbf{x}) \cdot\nabla
 \overline{P}(\mathbf{x})
+ (\beta k_i^2  - k_e^2)  E(\mathbf{x})   \overline{P}(\mathbf{x})
\right],
\label{dtempty}
\end{eqnarray}
where $E$ and $P$ are adequate forward and adjoint wave fields that solve transmission boundary value problems for stationary wave equations.
For biological samples $\beta=1$ and the gradient term disappears.
When no  initial guess for $\Omega$ is available, one sets $\Omega=\emptyset$ as initial guess, and computes the topological derivative when ${\cal R}=\mathbb{R}^3$. In this case,  $E(\mathbf x)=E_{inc}(\mathbf x)=e^{ik_e z}$ is the solution of the forward problem
\begin{eqnarray} \label{forwardempty}
\left\{
\begin{array}{l}
\Delta E + {k_e^2}E=0\quad\mbox{in ${\mathbb R}^3$},
\\[0.8ex]
\lim_{r\to \infty}r (\partial_r(E-E_{inc})-i{k_e}(E-E_{inc}))=0,
\end{array}
\right.
\end{eqnarray}
and
$\overline{P}({\mathbf x})
=- \sum_{j=1}^N {e^{ik_e |{\mathbf x}-{\mathbf x_j}|}
\over 4 \pi |{\mathbf x}-{\mathbf x_j}|}
(\overline{E_{meas,j}-E({\mathbf x_j})})$,  $P$ being the solution
of the adjoint problem:
\begin{eqnarray} \label{adjointempty}
\left\{
\begin{array}{l}
\Delta P + {k_e^2}P= \sum_{j=1}^N(E_{meas,j}-E) \delta_{{\mathbf x}_j}
\quad\mbox{in ${\mathbb R}^3$},
\\[0.8ex]
\lim_{r\to \infty}r (\partial_r P + i{k_e} P)=0,
\end{array}
\right.
\end{eqnarray}
$\delta_{\mathbf x_j}$ being Dirac masses supported at the detectors.

For small enough wavenumbers,  the largest negative values of the topological derivative are observed to concentrate inside the true objects, see \cite{carpiorapunln} and references therein.
Then, a first guess of the number and location of the scatterers may be obtained placing objects in regions where the topological derivative takes  the largest negative values.   Nevertheless, this strategy is only effective for small enough wavenumbers. As the wavenumber grows, the topological derivative develops oscillations. For single scatterers, and many incident directions,  the topological derivative is still useful since its  largest negative values concentrate around the object border \cite{carpiorapunln, guzinalarge}. For multiple objects, however, oscillations complicate the interpretation of the pattern \cite{carpiorapunln}.

An alternative concept has been developed to overcome these difficulties when the wavenumber increases. The topological energy is a useful complement  of the topological derivative, due to its ability to suppress oscillations. The topological energy is a scalar field defined as \cite{topologicalenergy}:
\begin{eqnarray}
E_T(\mathbf{x},{\cal R}) =  |E(\mathbf{x})|^2  |{P}(\mathbf{x})|^2,
\label{etempty}
\end{eqnarray}
where $E$ and $P$ are the forward and adjoint fields.  In particular, if ${\cal R}=\mathbb R^3$, then $E$ and $P$ are governed by (\ref{forwardempty})
and (\ref{adjointempty}) respectively.

\begin{figure}[]
\centering
\includegraphics[width=12cm]{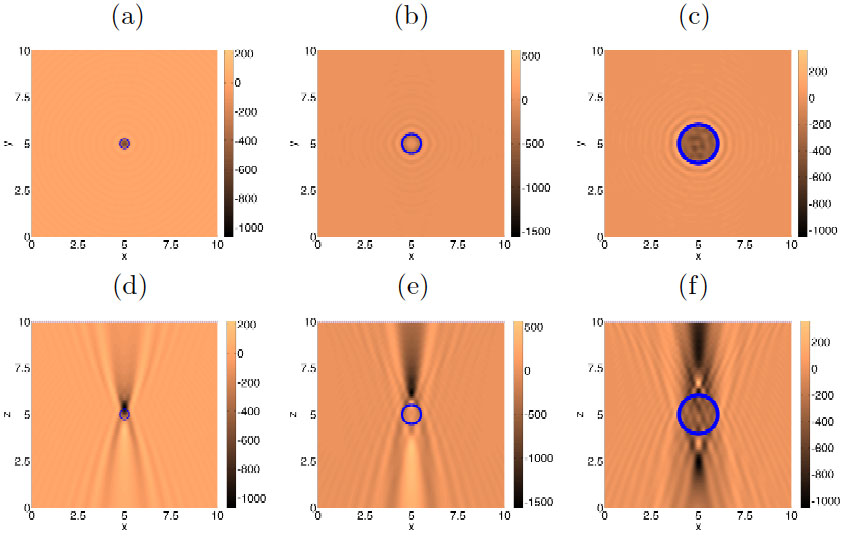}
\\
\includegraphics[width=12cm]{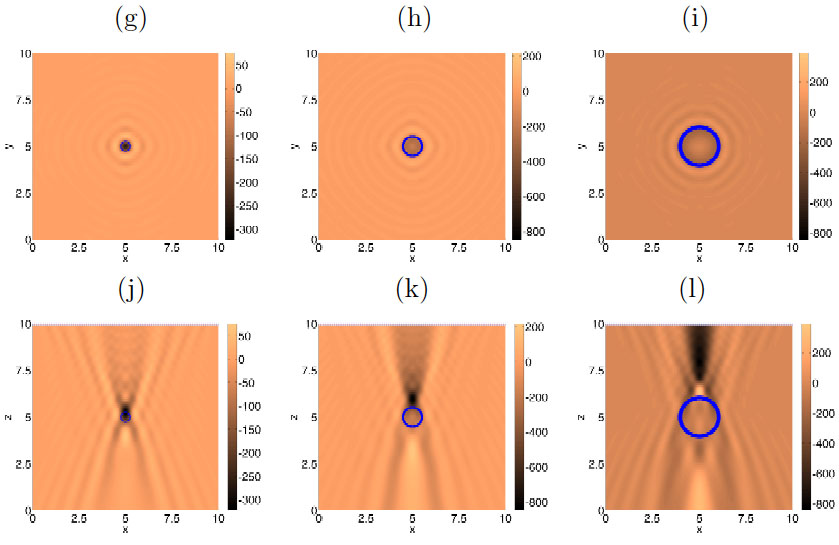}
\caption{
(a)-(c) Slices $z=5$ of the topological derivative evaluated with
Eqs. (\ref{dtempty}) and (\ref{forwardempty})-(\ref{adjointempty})  for
spheres of diameter $0.5$ ($0.5$ $\mu$m), $1$ ($1 \ \mu$m) and
$2$ ($2 \ \mu$m), respectively, centered at $(5,5,5)$ when
$k_e=20.6$, $k_i=24.79$ (violet light).
(d)-(f) Same for slices $y=5$.
(g)-(i) Same as (a)-(c) when $k_e=12.56$, $k_i=15.12$ (red light).
(j)-(l) Same as (d)-(f) when $k_e=12.56$, $k_i=15.12$.
The borderline of the true objects is superimposed.
}
\label{fig5}
\end{figure}

\begin{figure}[]
\centering
\includegraphics[width=12cm]{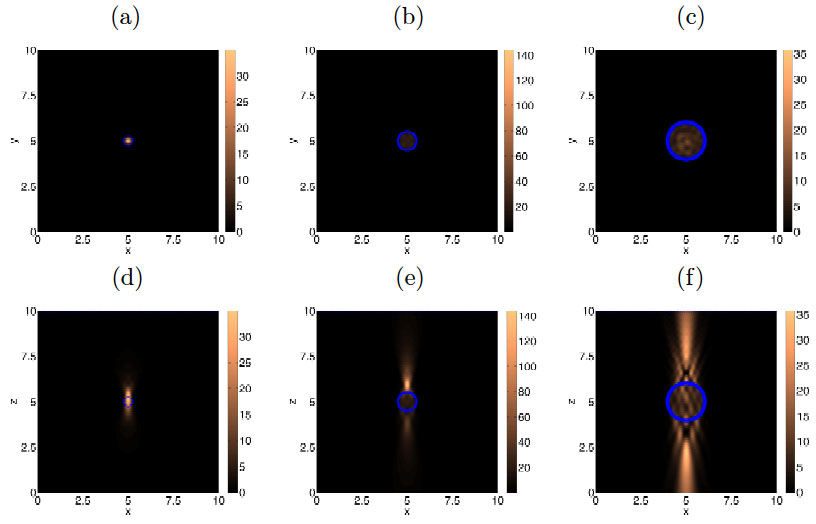}
\\
\includegraphics[width=12cm]{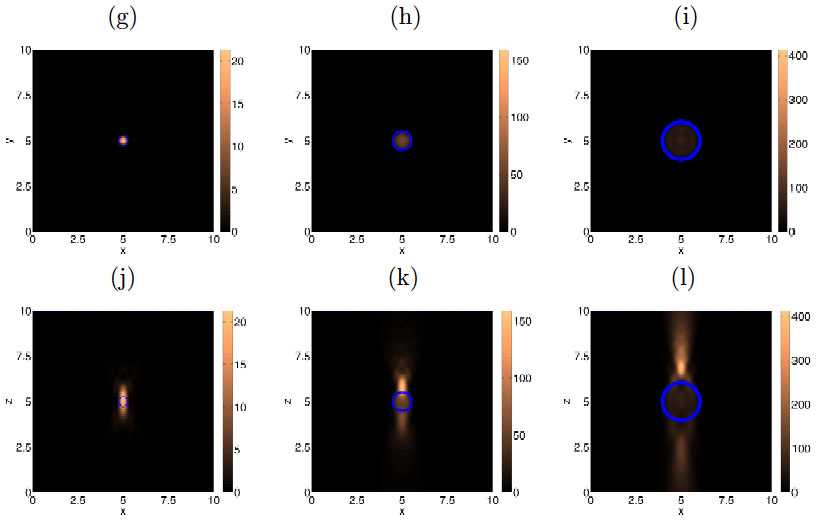}
\caption{
Counterpart of Figure \ref{fig5}, visualizing the topological energy evaluated with Eqs. (\ref{etempty}) and (\ref{forwardempty})-(\ref{adjointempty}), instead of the topological derivative.}
\label{fig6}
\end{figure}

Figures \ref{fig5} and \ref{fig6} compare the topological derivative and topological energy fields for balls with diameters  $0.5$ $\mu$m, $1$ $\mu$m and $2$ $\mu$m. In all cases, the $xy$ sections suggest the presence of balls of the right size and position in the $xy$ plane, as deduced from the concentration of large negative values of the topological derivative or high values of the topological energy in panels (a)-(c) and (g)-(i). The prediction of their distance to the detector screen, however, is more  uncertain and varies with the size and the employed wavelength, as observed in panels (d)-(f) and (j)-(l).  Indeed, along $xz$ or $yz$ slices, a peak shifted towards the screen is appreciable in most cases. As the size of the object increases and the wavelength decreases, we appreciate two peaks, placed before and after the true object   in the incidence direction. This fact is reminiscent of the above mentioned situation for increasing effective wavenumbers in two dimensions,  where the topological derivative concentrates around the whole boundary of the object when incident waves impinge from several incident directions around it.  Two-dimensional reductions focus precisely on axial studies  and show that, using one incident wave, the topological fields should attain their largest values before and after the object for large enough wave numbers. This fact may also be related to the analysis of caustics
performed in Ref. \cite{guzinalarge}. 

For our range of sizes, the topological fields permit to implement a systematic procedure to locate objects employing a single wavelength.
A simple strategy to define an initial guess $\Omega_0$ for the  scatterer sets
\begin{equation}
    \Omega_0:=\{{\bf x}\in {\mathbb R}^3 \,|\, E_T({\bf x},{\mathbb R}^3)\geq C_0\}
    \label{initialguess}
\end{equation}
where $C_0$ is a positive constant.  Alternatively, we may set
\begin{equation}
    \Omega_0:=\{{\bf x}\in {\mathbb R}^3 \,|\, D_T({\bf x},{\mathbb R}^3)\leq -C_0\}.
    \label{initialguesstd}
\end{equation}
In both cases, we check that $J({\mathbb R}^3 \setminus\overline{\Omega}_0)<J({\mathbb R}^3)$. Otherwise, we increase the constant $C_0$. 
 Notice that neither the definition of the topological energy
(\ref{etempty}) nor the analytical expressions for the forward and
adjoint fields (\ref{forwardempty})-(\ref{adjointempty}) depend on $k_i$.
Therefore, formula (\ref{initialguess}) provides a first approximation to
the objects without knowing their permittivities beforehand. Since 
the expression for the topological derivative with $\beta=1$ reduces to  
$D_T(\mathbf{x},{\cal R}) = {\rm Re}[
 (k_i^2  - k_e^2)  E(\mathbf{x})   \overline{P}(\mathbf{x})] $, the  
 function ${\rm Re} [E(\mathbf{x})   \overline{P}(\mathbf{x})]$ may 
yield a guess of the objects ignoring the variations of $k_i$ too.
We usually rely on topological energies to produce first guesses $\Omega_0$ since the images are often easier to interpret.  Nevertheless, for small enough isolated objects both procedures yield similar estimates and the quality of the predictions improves by iteration, see Sections \ref{sec:implementation} and \ref{sec:shape}. As we have commented above, a judicious interpretation of the topological fields may  yield better initial guesses of larger objects  selecting a range of $z$ between peaks of the topological fields and analyzing the $xy$ sections for that range of $z$. We will discuss this point in more detail in the next section.

If we wish to combine information from data measured for light of different wavelengths, we replace $E_T({\bf x},{\mathbb R}^3)$ in definition (\ref{initialguess})  by the normalized average of the topological energies computed for the two wavelengths at each point (each  of them is divided by its global maximum in the region ${\mathcal R}_{obs}$ under study): 
\[
E_{T,a}(\mathbf{x},{\mathbb R}^3) =
{1 \over M_1} E_{T,k_1}(\mathbf{x},{\mathbb R}^3)
+ {1 \over M_2} E_{T,k_2}(\mathbf{x},{\mathbb R}^3),
\]
where  $M_{\ell}= {\rm max}_{\mathbf{x}\in{\cal R}_{obs}} E_{T,k_{\ell}}(\mathbf{x},{\mathbb R}^3)$. A similar normalized average of the topological derivatives may be defined. However, we have not observed a noticeable improvement with these averaging procedures in our numerical tests,
though, for specific configurations one wavelength may perform better 
than the other.

Finally, if we had information from different incident directions, the topological energy fields would be just superimposed, see Section \ref{sec:shape}.

%
%

\section{Initial  three-dimensional reconstructions}
\label{sec:3D1st}

\begin{figure}[h!]
\centering
\includegraphics[width=10cm]{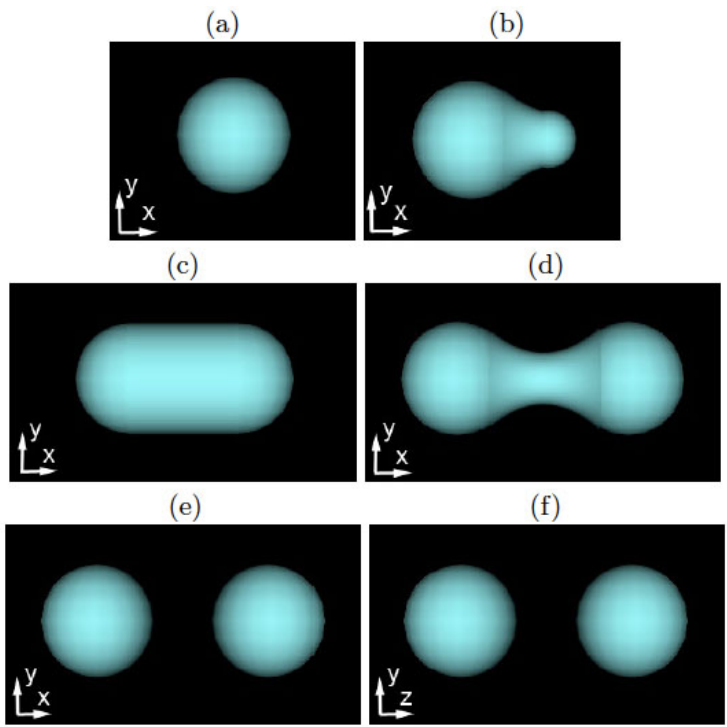}
\caption{ Three-dimensional objects used in the tests.
(a) sphere,
(b) pear-shaped object,
(c) spherocylinder,
(d) splitting spherocylinder,
(e) two spheres aligned parallel to the screen,
(f) two spheres aligned in the beam direction, orthogonal to the screen.}
\label{fig2}
\end{figure}

In this section we analyze the performance of  topological derivatives and energies  when visualizing biological objects of cell structure size using light beams. Figure \ref{fig2} displays a selection of symmetric, asymmetric, simply connected, multiply connected, convex and non-convex objects that we have considered in our reconstructions. We will see that these fields not only detect the presence of the defects, but are also able to locate them and to distinguish between different configurations and shapes. We can infer the structure of the objects from series of slices in the three space directions.

The imaging setting is detailed in Figure \ref{fig1} and Section \ref{sec:setting}.
After nondimensionalizing, the setting is as follows. We emit one incident wave in the direction $\mathbf{d}=(0,0,1)$, orthogonal to a planar screen located at the  plane $z=10$:
\[
 \Pi_{meas}=[0,10]\times[0,10]\times\{10\}.
\]
On this screen we place a uniform grid of detectors, using a step of 0.1 in 
the $x$ and $y$ directions, located at the points
 $\{(0.1k, 0.1\ell, 10)|$ $k,\ell=0,\dots,100\}.$
Typically, objects will be centered about (5,5,5) and we will compute the topological derivative/topological energy in the observation region
$
{\cal R}_{obs}=[0,10]\times[0,10]\times [0,10],
$
where the sample is located.
Topological derivatives and energies are computed evaluating formulas (\ref{dtempty}) and (\ref{etempty}), where the forward and adjoint fields are solutions of (\ref{forwardempty}) and (\ref{adjointempty}), respectively.
The data $E_{meas}$ are synthetically generated by solving the forward problem (\ref{forward}) for the true defect $\Omega$ using a coupled finite element-boundary element solver for Helmholtz equations inspired in mixed formulations employed in reference \cite{BEMFEM} for Laplace problems 
(for further details see also Section \ref{sec:implementation}).

For our numerical experiments we select two representative wavelengths, namely $\lambda^{\rm violet}=405\,  {\rm nm}$, $  \lambda^{\rm red}=660 \, {\rm nm}.$
The corresponding dimensionless parameters are given in Table \ref{table3}.
Notice that the assumption $\beta=1$ implies the disappearance of the gradient term in (\ref{dtempty}) when computing the topological derivative. In
the numerical experiments described here, the exterior and interior wavenumbers are $k_e=20.6$ and $k_i=24.79$ when using the violet light, or  $k_e=12.56$ and $k_i=15.12$ for the red one.  These wavenumbers correspond to typical values of lipid membranes surrounded by water. 
We have repeated the tests using the intermediate values for $k_i$ in Table \ref{table3}, obtaining similar results.  The values for $k_i$
are used to generate the synthetic data solving the forward problem and to
evaluate topological derivatives. The topological energy only depends on the given 
data and the value of $k_e$, but it does not depend on $k_i$.

\subsection{Influence of size and position}
\label{sec:size}

The structure of the topological derivative (TD) and topological energy (TE) fields changes when the size of the objects vary, as appreciated in Figures \ref{fig5} and \ref{fig6}.
For small objects, with diameters below a micron, the 3D regions where the topological derivative  takes large negative values or the topological energy attains its peaks mark the location of the object, giving a reasonable prediction of its size and shape, see panels (a), (d), (g), (j) in  Figures \ref{fig5} and \ref{fig6}. This prediction improves as the size of the object decreases.
Moreover, while the offset of the object center becomes negligeable using the largest wavelength (red light), the spurious elongation in the $z$ direction diminishes employing the shortest one (violet). In Section \ref{sec:shape} we 
will see that  the prediction of the position, shape and size 
of the obstacle may be improved by an iterative procedure.

On the contrary, for larger objects,  with diameters of a micron and above, the behavior of these fields varies from $xy$ to $xz$ or $yz$ slices.  Whereas large negative values of the TD and peaks of the TE reproduce the location, size and shape of $xy$ slices of the object (see panels (c),(i) in Figures
\ref{fig5} and \ref{fig6}), such values concentrate at the edges of the object in $xz$ or $yz$ slices.  In particular, panel (f) in Figures \ref{fig5} and \ref{fig6} locates the object between peaks of the TD and TE, providing a good guess of its thickness and position relative to the detector screen. This interpretation is clearer for the shorter wavelength in panel (f), than for the larger wavelength in panel (l).
A similar situation is observed for a smaller sphere in panels (b),(e),(h),(k) of Figures \ref{fig5} and  \ref{fig6}, with the largest values of the TD and TE located  past the object, shifted towards the detector screen, and a weaker peak before the ball.

The described behavior can be related to wavenumber variations observing that we nondimensionlize the problem using a reference object diameter $L= 1\,\mu$m. This yields the dimensionless wavenumber values (\ref{kL}) collected in Table \ref{table3}, that we fix for numerical purposes, while varying the object size. The effective dimensionless wavenumbers are obtained scaling (\ref{kL})  by a factor  ${{\cal L} \over L}$, where $\cal L$ is the diameter of the object under study. Their value decreases from panels in the center to panels in the left column of Figures \ref{fig5}-\ref{fig6} by a factor $2$ while it grows by a factor $2$ in the right column. As the size of the object diminishes, we enter 
the small effective wavenumber regime. 
Precision improves using the red light, that reduces the wavenumber, enhancing the concentration effect in the object occupied region.
Instead, by enlarging the object we enter a high effective wavenumber regime. Resolution improves employing the violet light, which increases the wavenumber further, enhancing the concentration at the borders in the axial direction. The transition to this regime is marked by a decrease in the magnitude of the TD and TE, consequence of the splitting of a peak in two.

Assuming impenetrable objects and short wavelengths, the theoretical analysis 
in Ref. \cite{guzinalarge} shows that the brightest 
points in TD/TE reconstructions are due to caustics. For impenetrable obstacles, 
the caustics appear exclusively on the back (i.e. `dark')
side of an obstacle.
Our objects being penetrable,  caustics may appear both at the back and 
the front sides of the anomaly. Secondary reflections off the `dark' side
of the boundary are possible, which would become more apparent
as the size of the object grows.
Our numerical simulations show that bright points on the front side 
become visible as the object size to wavelength ratio increases, 
and tend to be weaker than those on the back side, see Fig. 3. 
The threshold size might be characterized by the above mentioned
decrease in the magnitude of the TD.
Notice that whereas Ref. \cite{guzinalarge} takes measurements
on a sphere around the object, our measurements are made on a
screen past the object. This brings about additional factors, such
as the distance to the screen and the size of the screen relative
to the object, that influence the structure of the topological fields,
as we discuss in Section 4.4.

\begin{figure}[h!]
\centering
(a) \\
\includegraphics[width=8cm]{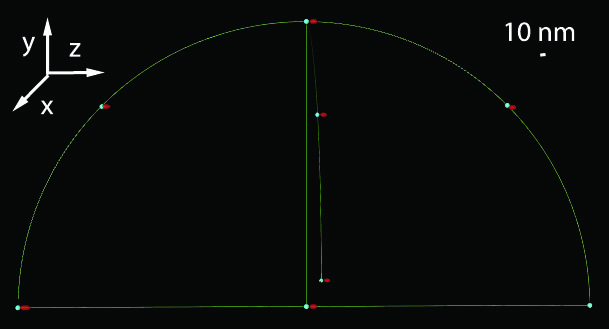} \\
(b) \\
\includegraphics[width=8cm]{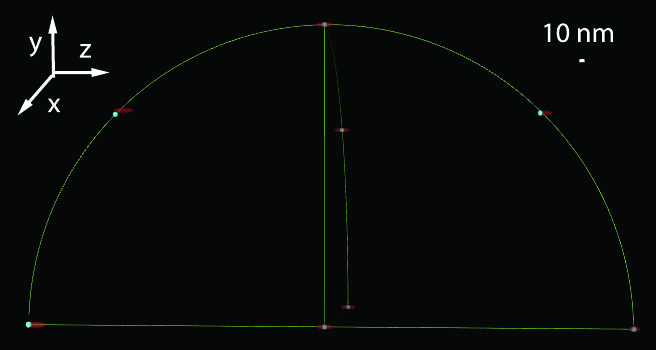}
\caption{
Superimposed reconstructions of  isolated balls of diameter $0.01$ 
($10$ nm) moving in circles of radius 1
($1 \ \mu$m) about the location $(5,5,5)$:
(a) Light of wavelength $405$ nm.
(b) Light of wavelength $660$ nm.
The true and reconstructed objects are represented in blue and red, respectively.}
\label{fig7}
\end{figure}

Let us analyze now the potential of the topological energy to locate the same defect at different positions.
These techniques could be employed to track the trajectories of nanoparticles, as evidenced by reconstructions with nanometer scale precision.
Figure \ref{fig7} compares the predicted and true objects, for isolated balls of diameter $0.01$ ($10$ nm) placed at different locations. Using the negative peaks of the TD defined
by formula  (\ref{initialguesstd}) with  $C_0=-0.9 \,{\rm min}_{{\bf x}\in \mathcal{R}_{obs}} \, D_T({\bf x},{\mathbb R}^3)$  we determine the center of the approximated objects, and the aspect ratio between the diameters in axial and lateral directions. Information about the true size is inferred from the magnitude of the TD and TE fields, that decreases noticeably with size. For $10$ nm objects, the TE takes maximum values of order $10^{-8}$ for the violet light and $10^{-9}$ for the red one. These values increase to  $30$ and $20$, respectively, considering  objects of $500$ nm diameter.
The reconstructed objects are ellipsoids, slightly elongated  in the direction $z$. Notice that the violet light produces less elongated guesses, whereas the red light determines position with more precision. For both wavelengths, the offset and the elongation decrease as we approach the screen.

\subsection{Shape identification}
\label{sec:shapeidentification}

\begin{figure}[h!]
\centering
\centering
\includegraphics[width=12cm]{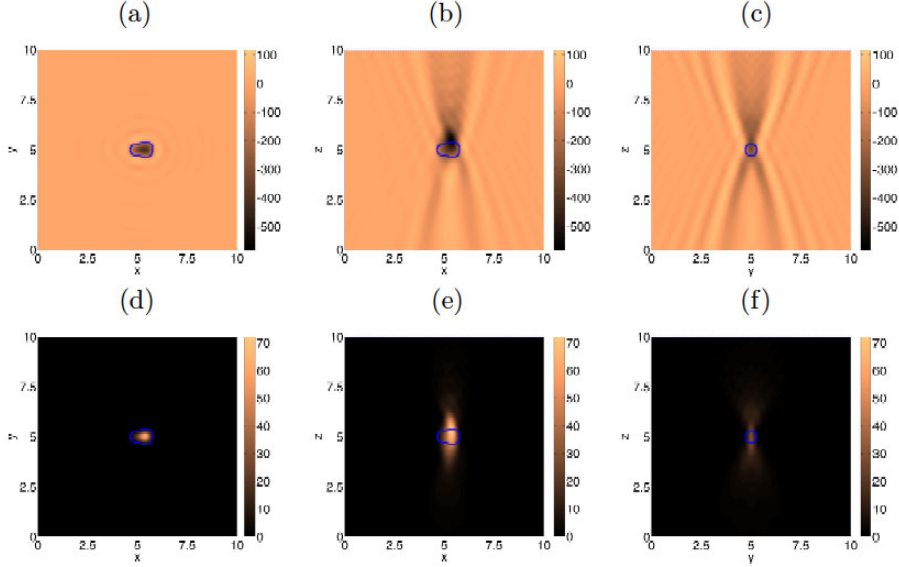}
\caption{Slices
(a) $z=5$, (b)  $y=5$  and (c) $x=5$ of the topological derivative
when $k_e = 12.56$ and $k_i = 15.12$ (red light)  for a pear-shaped
object of length $1$, and maximum diameter $0.5$.
(d)-(f) Same for the topological energy.
The blue borderline of the true object is superimposed.}
\label{fig8}
\end{figure}

\begin{figure}[h!]
\centering
\includegraphics[width=12cm]{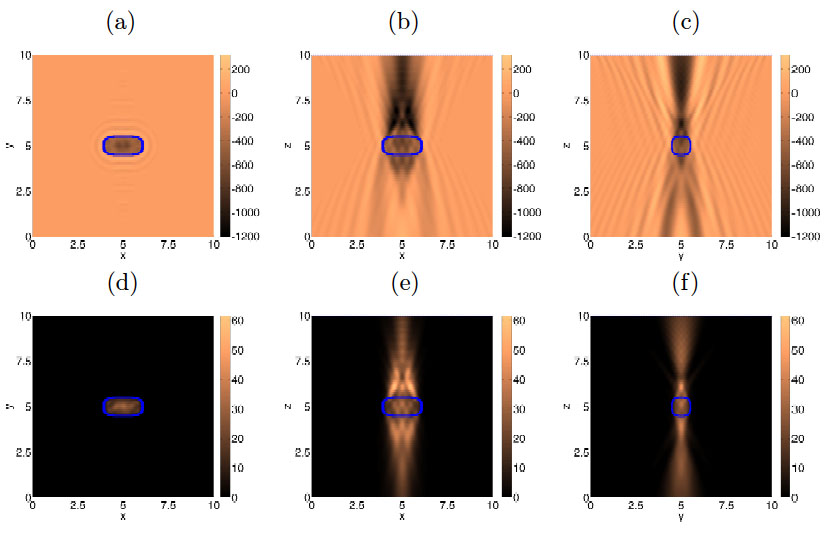}
\caption{Slices
(a) $z=5$, (b)  $y=5$ and (c) $x=5$ of the topological
derivative when $k_e = 20.6$ and $k_i = 24.79$ (violet light) for a
spherocylinder  of length $2$ and diameter $1$.
(d)-(f) Same for the topological energy.
The blue borderline of the true object is superimposed.
}
\label{fig3}
\end{figure}

\begin{figure}[]
\centering
\includegraphics[width=12cm]{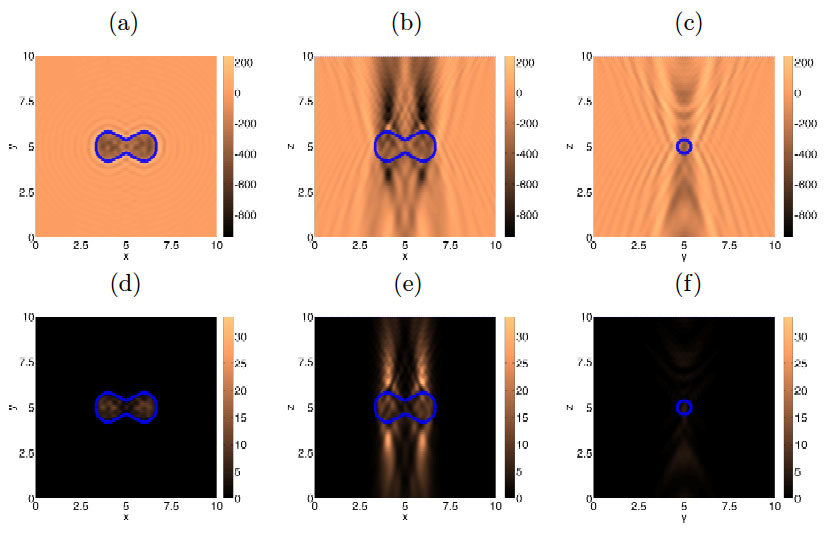}
\caption{ Slices
(a) $z=5$, (b)  $y=5$ and (c) $x=5$ of the topological
derivative when $k_e = 20.6$ and $k_i = 24.79$ (violet light) for a
splitting spherocylinder of length $2$ and maximum diameter $1$.
(d)-(f) Same for the topological energy.
The blue borderline of the true object is superimposed.}
\label{fig9}
\end{figure}

\begin{figure}[]
\centering
\hskip -1cm (a) \hskip 5cm (b) \\
\includegraphics[height=5cm]{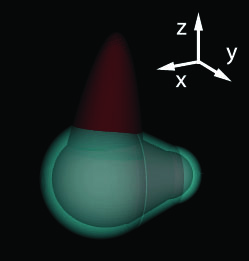}
\includegraphics[height=5cm]{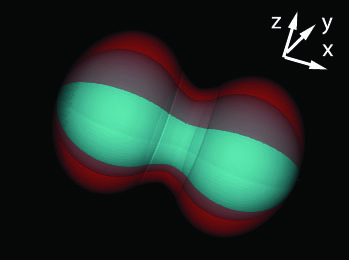}
\caption{ (a) Possible initial reconstruction (red) of the pear-shaped
object of length $1$ and maximum diameter $0.5$ using a light  beam
of wavelength $660$ nm, orthogonal to the detector screen.
(b) Possible initial reconstruction (red) of the splitting rod
of length $2$ and maximum diameter $1$ using a light  beam of
wavelength $405$ nm.}
\label{fig4}
\end{figure}

Precise information about the shape of the objects can be inferred from $xy$ slices of TD and TE fields. Comparing panels (a),(g) in  Figures \ref{fig5} and \ref{fig6} to Figure \ref{fig8}(a) and (d) we distinguish the round shape corresponding to a sphere from the pear-like structure depicted in Figure \ref{fig2}(b).
Figure \ref{fig3} represents topological energies  and derivatives for the rod-like object shown in Figure \ref{fig2}(c). A narrower central part is appreciated in Figure \ref{fig9}(a) 
and (d), suggesting a splitting rod, the sand clock shaped object depicted in Figure \ref{fig2}(d). Provided the distance of the object to the screen is approximately known, the slices $z=z_0$ of the TD and TE fields allow us to identify the shape of the object, and track changes in its appearance, including asymmetric and non-convex forms.

Besides, the approximate location in the direction $z$ is deduced from the $xz$ and $yz$ slices.  Indeed, the prediction of the position is good for the
pear-like object using light of wavelength $660$ nm, where the range $z_0$ of interest corresponds to the peak of the TE. The prediction of the position obtained in this way improves for smaller objects.  On the contrary, for the slightly larger  rod and the splitting rod, we may use the $405$ nm light. Then, the range of interest for $z_0$ is comprised between the two peaks of the TE. Gathering the information provided by the slices $z=z_0$ we propose the reconstructions shown in Figure \ref{fig4}.
Panel (a)   in this figure uses values $z_0$ corresponding to the only peak, whereas panel (b) selects values of $z_0$ between the two peaks.
More precisely, the reconstruction procedure for a single object is:
\begin{itemize}
\item Decide if the effective wavenumber is such that: (a) single objects are located at peaks, or (b) single objects are located between peaks of the TE.
\item Determine the relevant range $[z_1,z_2]$ of $z$ to be considered:
\begin{itemize}
\item In case (a), $[z_1,z_2]$ is determined by the lower and upper limits in the $z$ direction of the peak representing the object. Such peak is defined by formula (\ref{initialguess}),
where $C_0$ is given by
\begin{equation}\label{c0}
C_0=\alpha_0\max_{{\bf x}\in\mathcal{R}_{obs}}E_T(\mathbf x,\mathbb R^3),
\quad 0 < \alpha_0 < 1,
\end{equation}
and $\alpha_0$ is a value to be calibrated.  For example, for the peak-like object we may take $\alpha_0=0.5$.
\item In case (b), the range $[z_1,z_2]$ of $z$  is determined by the location of the maxima of the upper and lower peaks.
\end{itemize}
\item  For $z_0 \in [z_1,z_2]$, study the slices $z=z_0$ of the topological fields. The rings about the region where the TD takes the largest negative values mark its border. The $z=z_0$ slice of the object may be estimated from
\begin{equation}\label{c0bis}
\left\{ \mathbf{x}=(x,y,z_0) \big| \, D_T(\mathbf x,\mathbb R^3) < -C_{z_0}
\right\},
\end{equation}
where $C_{z_0}=-\alpha_0\min_{{\bf x}\in {\cal R}_{obs}\cap \{z=z_0\}}D_T({\bf x},\mathbb R^3)$.
\end{itemize}
The  three-dimensional images in Figure \ref{fig4} are VRML (virtual reality modeling language) representations of simple surfaces approximating the reconstructed objects.

\subsection{Multiple objects}

Standard vision does not detect objects placed behind other objects along the visual line. Moreover, it sometimes identifies objects close to each other as if they were just a single object. On the contrary, our method allows us to distinguish scatterers that would in principle be hidden by previous obstacles in the direction of the incident light; and it is also able to distinguish between multiple defects that are very close. 

\begin{figure}[h!]
\centering
\includegraphics[width=12cm]{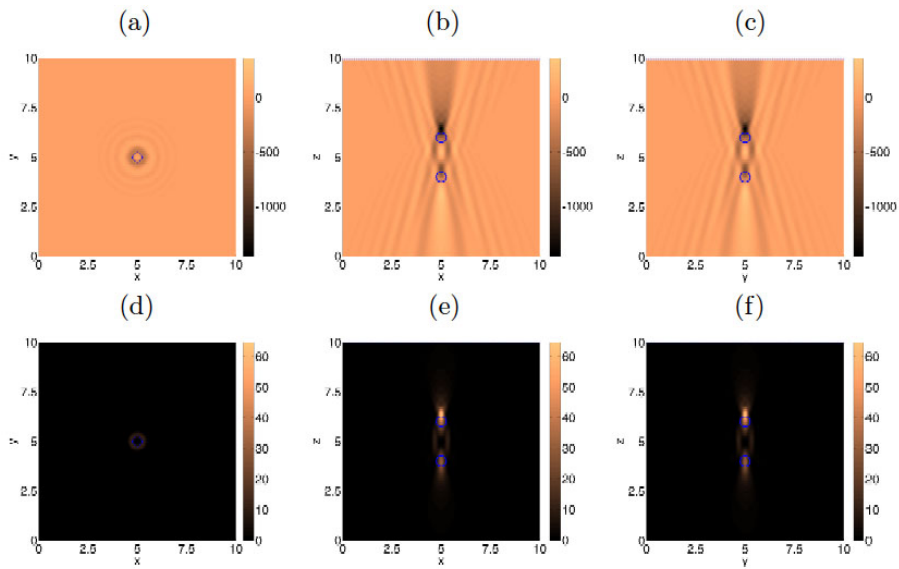}
\caption{ Slices
(a) $z=5$, (b)  $y=5$ and (c) $x=5$ of the topological derivative when $k_e = 20.6$ and $k_i = 24.79$ (violet light). The true objects are two spheres of diameter $0.5$, situated one
behind the other along the axis $z$, at distances $4$ and $6$ from the screen.
(d) Same for the topological energies. The contour of the true objects is superimposed in (b)-(c) and (e)-(f). No true object is present in (a),(d), though we have superimposed a ghost contour of the balls as reference.}
\label{fig10}
\end{figure}

To illustrate the ability of our method to distinguish objects placed along the same incident direction, Figure \ref{fig10} analyzes two identical spheres placed along the $z$ axis, as in Figure \ref{fig2}(f). They can be identified with both the $405$ nm and  $660$ nm lights.  Indeed, unlike the single ball considered in panels (c) and (f) of Figures \ref{fig5} and \ref{fig6}, the hollow  $xy$ slices between the two peaks indicate the presence of two objects placed at the peaks, being the object situated closer to the detector screen the most prominent.

\begin{figure}[h!]
\centering
\includegraphics[width=12cm]{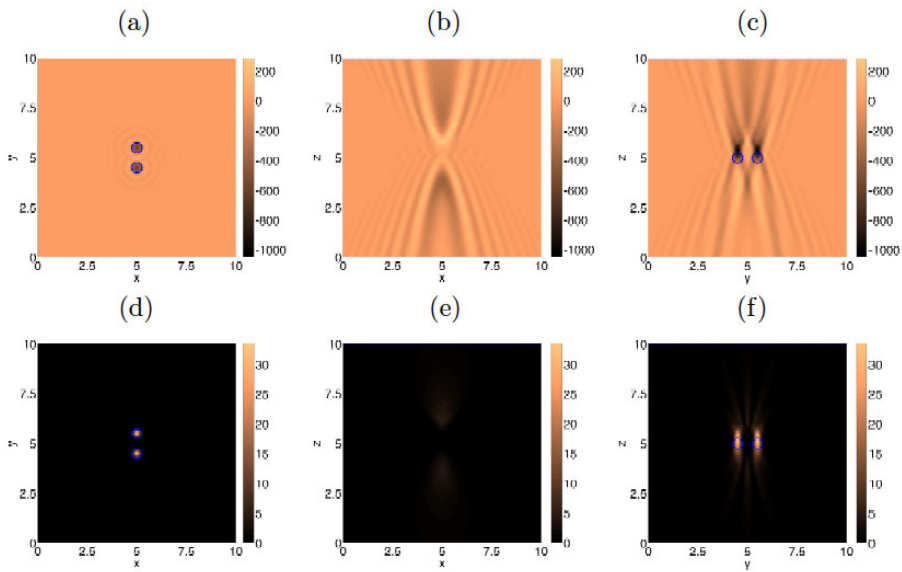}
\caption{
Slices (a) $z=5$, (b)  $y=5$  and (c) $x=5$ of the topological derivative when  $k_e=20.6$ and $k_i=24.79$ (violet light). The true objects are two spheres of diameter $0.5$, separated a distance $0.5$ on the plane $z=5$.
(d)-(f) Same for the topological energies. The contour of the true objects is superimposed in (a),(c),(d) and (f).
}
\label{fig11}
\end{figure}

Additional tests show that close objects are distinguishable. Figure \ref{fig11} represents two spheres located on the plane $z=5$, along the $y$ axis. They are perfectly identified with  both wavelengths, though the shortest one yields better resolved shapes. In both cases, the intermediate gap is completely evident, without any blurring. Similar results are observed slightly increasing or decreasing the distance between the objects.

Repeating these tests with objects characterized by different
values of $k_i$, we obtain similar results, except that the objects with higher
contrast $k_e^2-k_i^2$ are more clearly distinguished. Notice that topological
energies are computed without making use of $k_i$. When these values
are unknown, the topological energies  would yield information on the relative magnitudes of $k_i$ for different objects that might be improved by gradient methods  \cite{carpiorapun1,carpiorapun2}.
For an increasing number of close spheres, objects blur as the distance
between the spheres diminishes, specially in the direction of the incident
waves. Resolution is similar provided they are well separated. The topological
fields used to define first approximations of the objects use analytical
formulas, therefore the number of objects does not pose a difficulty at this
stage. It may raise the computational cost of later iterative corrections
relying on BEM-FEM methods to evaluate forward and adjoint fields, though.

\subsection{Influence of the screen size and other parameters}

After performing series of numerical experiments varying some of the parameters involved in the imaging setting, we have extracted the following conclusions:

\begin{itemize}  
\item Fixing the screen size and the distance between detectors, reconstructions worsen when we move the object away from the screen or we displace it left or right noticeably.

\item For a fixed screen size,  once enough detectors have been placed, increasing the number does not provide significant improvements (replacing the detectors $\{(0.1k, 0.1\ell, 10) | \, k,\ell=0,\dots,100\}$ by $\{(0.05k, 0.05\ell, 10)| \, k,\ell=0,\dots,200\}$, for instance).

\item For a fixed object size, precision and resolution worsen or improve noticeably by reducing or enlarging the size of the screen. This is specially true for the predicted distance to the screen,  in examples such as Fig. \ref{fig5}-\ref{fig6} (d).

\end{itemize}

We have repeated some of the previous experiments considering the material parameters for nucleus and organelles instead of those for lipid membranes (see Table \ref{table3}), obtaining similar results. We have also observed that the method is not too sensitive to noise.

\section{Postprocessing by iteration}
\label{sec:implementation}

An inspection of the topological energy and derivative of the cost functional (\ref{cost}) when $\Omega= \emptyset$ provides a first guess of the number, location and size of the objects,
by means of analytical formulas involving the data, the incident
wave and the value of $k_e$ for the background medium.
If we are only interested in tracking the location or the displacement of objects, or changes of shape in $xy$ sections, this stage may suffice for simple configurations. In general, there is still incertitude about the true number, size, shape and location of objects.  Developing methods that are able to improve a first guess $\Omega_0$ using only the stored measured data to produce more precise information about the structure of the scatterers $\Omega$ is essential for applications. Here, we introduce a  three-dimensional  framework to implement topological derivative and energy based iterative schemes overcoming computational difficulties. We set $\beta=1$  and assume $k_i$ known, that is often the case when
we are just tracking positions and changes of shape.
In a more general situation, one can search for both the  object $\Omega$ and their material parameters ($\varepsilon_i$ and $\mu_i$ or equivalently, $\beta$ and $k_i$ in this framework). In this case, gradient techniques might yield complementary information about them \cite{carpiorapun1,carpiorapun2,carpiorapun3}.

\subsection{Topological energy based reconstruction  algorithm}

A simple strategy to define an initial guess $\Omega_0$ for the scatterer sets was proposed in Section \ref{sec:topological} through formula (\ref{initialguess}). The positive constant $C_0$ is defined by (\ref{c0}), with $0<\alpha_0<1$. We have used $\alpha_0=0.75$ in our iterative  tests.

Once a guess for the objects $\Omega_0$ is available, we can modify  it evaluating an updated topological energy field. A possible iterative scheme is the following: \\

\begin{itemize}
\item {\em Initialization:} Define $\Omega_0$ by
formula (\ref{initialguess}).

\item {\em Iteration:} At each step, a new object $\Omega_{n+1}$ is constructed from $\Omega_{n}$ setting
\begin{equation}
    \Omega_{n+1}:=\{{\bf x}\in  \mathbb R^3
    \,|\, E_T({\bf x},{\mathbb R}^3\setminus
    \overline{\Omega}_{n})\geq C_{n+1}\},
    \label{updatedguess2}
\end{equation}
where the topological energy is given by:
 \begin{eqnarray}
E_T(\mathbf{x},{\mathbb R}^3 \setminus
    \overline{\Omega}_{n}) =  |E_n(\mathbf{x})|^2  |
    {P}_n(\mathbf{x})|^2, \quad{\bf x}\in\mathbb R^3,
\label{etfull}
\end{eqnarray}
$E_n$ and $P_n$  being forward and adjoint fields with object $\Omega_n$. 
This is to say, $E_n$ is the solution of the forward problem:
\begin{equation}\label{forwardEnergy}
\left\{\begin{array}{ll}
\Delta E_n+k_e^2 E_n=0,&\quad\mbox{in $\mathbb R^3\setminus\overline{\Omega}_n$},\\
\Delta E_n+k_i^2 E_n=0,&\quad\mbox{in $\Omega_n$},\\
E_n^-= E_n^+,&\quad \mbox{on $\partial\Omega_n$},\\
\beta \partial_{\mathbf{n}}  E_n^-= \partial_{\mathbf{n}} E_n^+, &\quad 
\mbox{on $\partial\Omega_n$},\\
\displaystyle\lim_{r\to \infty} r\big(\partial_r (E_n- E_{inc})-\imath k_e(E_n- E_{inc})\big)=0,
\end{array}
\right.
\end{equation}
and  $P_n$ is the solution of the adjoint problem:
\begin{eqnarray} \label{adjoint}
\left\{
\begin{array}{ll}
\Delta P_n + {k_e^2}P_n= \sum_{j=1}^N(E_{meas,j}-E_n(\mathbf{x}_j)) \delta_{{\mathbf x}_j}&\quad\mbox{in ${\mathbb R}^3\setminus\overline{\Omega}_n$},
\\
\Delta P_n+k_i^2 P_n=0,&\quad\mbox{in $\Omega_n$},\\
P_n^-= P_n^+,&\quad \mbox{on $\partial\Omega_n$},\\
\beta \partial_{\mathbf{n}}  P_n^-= \partial_{\mathbf{n}} P_n^+, &\quad \mbox{on $\partial\Omega_n$},\\
\displaystyle\lim_{r\to \infty} r\big(\partial_r P_n
+\imath k_eP_n\big)=0.
\end{array}
\right.
\end{eqnarray}
 In practice, we compute the auxiliary field $\overline{P}_n$, which is the solution of a problem with the same structure as the forward problem, setting the incident field equal to zero and adding a source $\sum_{j=1}^N(\overline{E_{meas,j}-E_n(\mathbf{x}_j)}) \delta_{{\mathbf x}_j}$. Notice that these forward and adjoint fields are defined everywhere, as well as the topological energy.

\item {\em  Stopping criteria:}
The iteration should stop if either meas$(\Omega_{n+1}\setminus\Omega_{n})$ is small enough, or $J(\mathbb R^3 \setminus\overline{\Omega}_{n+1})$ is small enough, or the difference of the functional values between consecutive steps is small enough,  or the discrepancy principle $\frac1N\sum_{j=1}^N|E_{meas,j}^\varepsilon-E_{n+1}(\mathbf{x}_j)|<\tau\varepsilon$
is satisfied. Here, $\varepsilon$ describes measurement errors and $\tau>1$.
\end{itemize}

Alternative strategies that only add points to existing guesses forbid removing spurious points, whereas our technique redefines the whole approximation again, allowing us to discard spurious points.
For large wavenumbers our method may produce improved guesses of the objects, though convergence is uncertain.

\subsection{Topological derivative based reconstruction algorithm}

An initial guess $\Omega_0$ for the scatterers set was proposed in Section \ref{sec:topological} through formula (\ref{initialguesstd}).  
We may take
\begin{equation}\label{c0td}
C_0=- \alpha_0 \min_{{\bf x}\in\mathcal{R}_{obs}}D_T(\mathbf x,\mathbb R^3),
\end{equation}
with $0<\alpha_0<1$. We have chosen  $\alpha_0=0.75$ in our iterative tests.

The guess $\Omega_0$ can be improved updating the topological derivative field to take into account its presence. The following  procedure  has been used for two-dimensional tests in acoustics \cite{carpiorapun1}. We adapt it here to our  three-dimensional holographic setting.\\

\begin{itemize}
\item {\em Initialization:} Define $\Omega_0$ by formula (\ref{initialguesstd}).
\item {\em Iteration:} At each step, a new object $\Omega_{n+1}$ is constructed from $\Omega_{n}$ setting
\begin{eqnarray}
    \Omega_{n+1}:= \{{\bf x}\in \Omega _n  \,|\, D_T({\bf x},{\mathbb R}^3\setminus
    \overline{\Omega}_{n})< c_{n+1}\}
    \nonumber\\
\hspace*{1cm} \cup \, \{{\bf x}\in  \mathbb R^3 \setminus\overline{\Omega}_n    \,|\, D_T({\bf x},{\mathbb R}^3\setminus \overline{\Omega}_{n})<- C_{n+1}\}.
    \label{updatedguess2td}
\end{eqnarray}
For the iterations with the pear-like object discussed in Section \ref{sec:shape}, we set
\begin{eqnarray}
C_1= -\alpha_1 \min_{{\bf x}\in\mathcal{R}_{obs}}D_T(\mathbf x,\mathbb R^3),
\quad
c_1=\beta_1 \max_{{\bf x}\in\mathcal{R}_{obs}}D_T(\mathbf x,\mathbb R^3),
\label{cC1}
\end{eqnarray}
with $\alpha_1=1/2$ and $\beta_1=1/80.$

Notice that, at each stage, points where the topological derivative is negative and large are added to existing objects, whereas points where the topological derivative is positive and large are removed from existing objects.
The topological derivative is defined for all ${\bf x}\in \mathbb R^3$ by \footnote{This formula was employed in the 2D numerical tests for acoustics performed in \cite{carpiorapun1}. There is a sign mismatch in that paper between the theoretical definition given for the  TD inside the object and the definition used in the codes for the numerical tests,  that is the one with this practical interpretation.}:
\begin{eqnarray}
D_T(\mathbf{x},{\mathbb R}^3 \setminus
    \overline{\Omega}_{n}) = {\rm Re} \Big[ 
(k_i^2  - k_e^2)  E_n(\mathbf{x})   \overline{P}_n(\mathbf{x})
\Big], \label{dtfull}
\end{eqnarray}
$E_n$ and $P_n$  being forward and adjoint fields with object $\Omega_n$, governed by systems (\ref{forwardEnergy}) and (\ref{adjoint})  respectively.
The positive constants $C_{n+1}$, $c_{n+1}$ in (\ref{updatedguess2td}) are  ultimately chosen to ensure decrease of the cost functional, i.e., $J({\mathbb R}^3 \setminus\overline{\Omega}_{n+1})<J({\mathbb R}^3 \setminus\overline{\Omega}_{n})$. Initially, we may set $C_{n+1}=C_n$ and $c_{n+1}=c_n$. If the restriction is violated, the values are increased.

\item {\em Stopping criteria:}  The same as for the topological energy based method.
\end{itemize}

For small enough effective wavenumbers, topological derivative based schemes are expected to produce sequences of approximated objects decreasing the cost functional in view of expansion (\ref{dteout}). The effective wavenumbers must be such that the object location in the
$z$ direction is marked by one peak about the object location,
not two at its edges, as discussed in Section \ref{sec:size}.

\subsection{Implementation}

A few technical details about the implementation are in order. The tests presented in Sections \ref{sec:3D1st} and  \ref{sec:shape} use synthetic data. We place the true objects between the incident wave and the detector screen in the setting described in Sections \ref{sec:intro} and \ref{sec:setting}. Then, we compute the synthetic data solving the forward problem and evaluating the electric field at the detectors. Known the data, we implement the algorithms described in Section \ref{sec:topological} and in this section to reconstruct the objects.

Each object $\Omega$ is defined using a signed distance function $f_d$, negative inside $\Omega$ and positive outside:
\begin{equation}
\Omega= \{ {\mathbf x} \in \mathbb R^3 \big|\,  f_d({\mathbf x})<0 \}.
\label{distance}
\end{equation}

To solve the forward problem (\ref{forward}) we mesh the object and use a non-symmetric  BEM-FEM scheme.
More precisely, we generate an unstructured mesh on the object using the open access mesher {\em DistMesh}  described in Ref. \cite{mesh}.  
For the discretization spaces, we use continuous and piecewise $\mathbb{P}1$ finite elements, as well as piecewise constant  boundary elements. The boundary integrals entering both the BEM formulation and the BEM-FEM coupling integrals are evaluated following \cite{integrals}. In particular, boundary integrals are approximated using the Duffy transform, which is a local change of variables that regularizes possible singularities allowing for standard numerical integration (for example, Gauss formulas with six nodes in each direction in our tests). It is well known that the resulting method has order one.
We have selected this discretization because it avoids domain truncation or infinite elements. Moreover, it is  flexible enough to deal with a wide range of shapes. When $k_e$ and $k_i$ are constant, faster pure BEM-BEM methods \cite{bruno} could be used. The BEM-FEM strategy is well suited for  more general situations that include the presence of  regions with variable parameters inside the objects, or for future combination with gradient strategies to invert  also the interior parameters.

The approximated solution of the forward problem for the true objects is evaluated at $101 \times 101 = 10201$ detectors on $[0,10]\times [0,10] \times \{10\}$, on a grid with $101$ values uniformly distributed in each direction:
\begin{equation}
\mathbf{x}_{k,\ell}= (x_{k},y_{\ell},10) = (0.1k,0.1\ell,10), \label{receptors}
\end{equation}
with $k,\ell=0,...,100$. For each detector, we compute an approximation of the measured wave $E_{meas,k,\ell}\approx E(\mathbf{x}_{k,\ell})$ (where $E$ is the solution of the forward problem (\ref{forward}) for the true defects) using an integral representation of the solution of the scattering problem. More precisely, the solution of the BEM-FEM scheme provides us with approximations of the Cauchy data of the solution of the forward problem, which we can use in Green's formula and then integrate numerically on each triangle on the boundary (for example, with the barycenter formula in our tests). Further, if necessary, we may compute partial derivatives of the solution just differentiating the integral representation and proceeding as before.  

Once an approximation of the measured data is available, we estimate the initial value of the cost functional
\begin{equation}
J({\mathbb R}^3) =\frac12  \sum_{k,\ell=0}^{100} |E_{inc}(\mathbf{x}_{k,\ell})-E_{meas,k,\ell}|^2.
\label{initialcost}
\end{equation}

The first topological derivatives and energies are computed using formulas (\ref{dtempty}) and (\ref{etempty}) with forward and adjoint fields given by (\ref{forwardempty}) and (\ref{adjointempty}), respectively.
We generate a uniform grid of sampling points  $\{\mathbf{y}_1,\dots, \mathbf{y}_{N_{obs}}\}$ in the region we want to observe, ${\cal R}_{obs}$. In our numerical tests, ${\cal R}_{obs}=[0,10]\times[0,10]\times[0,10]$, and the uniform grid contains $N_{obs}=8\times 10^6$ points, obtained by considering 200 points in the $x, y$ and $z$ directions.

To apply the iterative method based on successive computations of topological energies (\ref{etfull}) and derivatives (\ref{dtfull}), we need to define $\Omega_0$ (and the subsequent approximations $\Omega_m$) 
by means of a signed distance function which is smooth to be properly meshed with \emph{DistMesh}. 
Besides, we want the initial objects $\Omega_0$ to approximate identities (\ref{initialguess}) or (\ref{initialguesstd}), whereas the successive corrections $\Omega_m$ must approximate the selected definition (\ref{updatedguess2}) or (\ref{updatedguess2td}).
Notice that we need to mesh $\Omega_m$ to evaluate the cost functional $J({\mathbb R}^3\setminus\overline{\Omega}_m)$ corresponding to the proposed reconstruction using formula (\ref{cost})
and to compute adjoint and forward fields in the next iteration. In order to define an adequate signed distance function $f_d$   for expression (\ref{distance})  to characterize these domains when meshing with \emph{DistMesh}, we proceed in several stages:
\begin{enumerate}
\item  We select a small box $[m_1,M_1] \times [m_2,M_2] \times [m_3,M_3]$ containing $\Omega_m$, and we generate a uniform grid of points $\{\mathbf{q}_1,\dots,\mathbf{q}_S\}$, with a small enough step in each direction $d_q$ to ensure that a sufficient number of points are inside $\Omega_m$ (for example,  
in our experiments we have taken about $50-110$ auxiliary points inside).

Notice that this step could be avoided by considering a finer grid $\{\mathbf{y}_1,\dots, $ $\mathbf{y}_{N_{obs}}\}$ from the beginning.  
However, if such grid is very fine, then the computational cost in time and memory becomes unnecessarily high. For this reason, we use a somehow coarse grid of the observation region to localize the possible region where the object is located, and once this region has been identified, we consider  a finer grid only in that region.

\item We approximate the topological energy in the fine grid $ \{\mathbf{q}_1,\dots,\mathbf{q}_S\}$ by linear interpolation of the values in the coarse grid $\{\mathbf{y}_1,\dots,  \mathbf{y}_{N_{obs}}\}$.

This is faster than using the direct evaluation of the adjoint and forward fields through Green's formula and, moreover, it smooths the topological fields.

\item We select from  $\{\mathbf{q}_1,\dots,\mathbf{q}_S\}$ the points in which the topological energy takes values
above thresholds and denote them $\{\widetilde{\mathbf{q}}_1,\dots,\widetilde{\mathbf{q}}_{S'}\}$.

\item We define $\Omega_{m}$  as the smooth union of balls of radius $d$ centered at the points $ \{\widetilde{\mathbf{q}}_1,\dots,\widetilde{\mathbf{q}}_{S'}\}$ following the blobby molecule strategy introduced in \cite{blobby}. More precisely, at any point $ \mathbf{x}$ the signed distance function associated to $\Omega_m$ is defined as:
\begin{equation}\label{fd}
f_{d,\Omega_m} (\mathbf{x}) = \exp(-1) - \sum_{j=1}^{S'} \exp(-|\mathbf{x}-\widetilde{\bf q}_j|^2/(\alpha\,d_q^2)), \label{distancem}
\end{equation}
Notice that the radius of each blobbly molecule is $d=d_q \sqrt{\alpha}$, $d_q$ previously selected at step 1. We may set $\alpha=1$, but it can be increased, for instance to $\alpha= 2$; increasing the value of $\alpha$ we obtain a smoother domain, but the size of the domain also increases.

\item Once $\Omega_{m}$ is defined, we generate a mesh using {\em DistMesh} and check the mesh quality and the number of nodes. If the values are reasonable, we keep the mesh. Otherwise, when the quality of the mesh is inadequate we modify the points $\{\mathbf{q}_j\}$ (that is, $d_q$) or the parameter $\alpha$. Besides, if the problem is related to the number of nodes, we modify the diameter of the mesh of $\Omega_{m}$.

\end{enumerate}

\section{Shape corrections}
\label{sec:shape}

Once we have a guess of the number of scatterers, their location and size, we can improve it using the same set of data. Notice that the achievable resolution is  limited by the choice of incident directions. 
Moreover,  once enough detectors have been distributed, rising the number  does not enhance resolution, though increasing the size of the screen may sharpen it.  Let us test how the postprocessing algorithms described in Section \ref{sec:implementation} may be exploited to refine the first guesses constructed in Section \ref{sec:3D1st}.

\begin{figure}[h!]
\centering
\includegraphics[width=12cm]{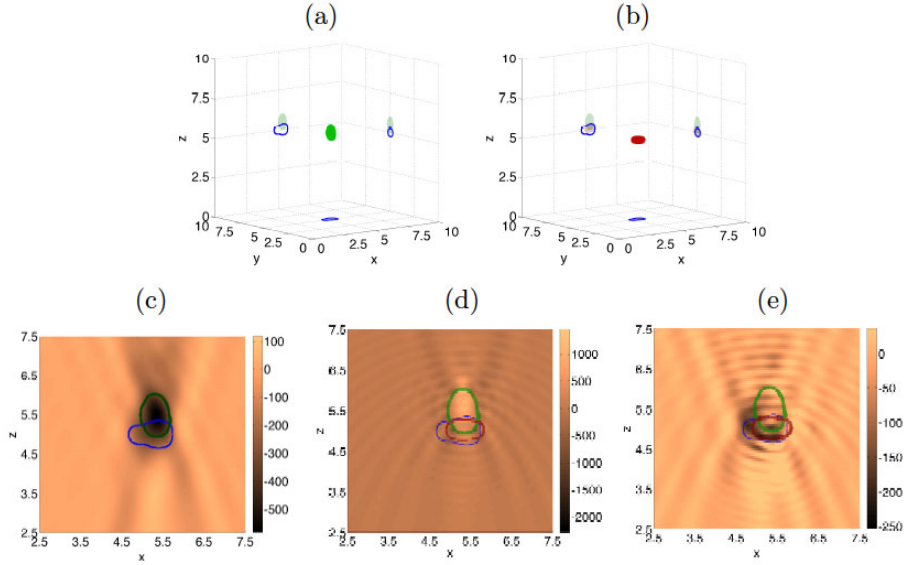}
\caption{
Succesive approximations of the pear-like object in Figure
\ref{fig8} constructed from updated computations of topological
derivatives: (a) first guess $\Omega_0$, (b) second
reconstruction $\Omega_1$.
Projections on the coordinate planes
compare the projection of the true object (blue line) with the
projected reconstructions (solid green for $\Omega_0$,
solid pink for $\Omega_1$).
Slices  $y=5$ of the topological derivative for the
first, second and third reconstruction are shown in (c)-(e).
The last one suggests adding and removing points to approach
the pear-like shape. The blue contour represents
the true object, the green contour the first guess,
the red contour the second.}
\label{fig12}
\end{figure}

\begin{figure}[h!]
\centering
\includegraphics[width=12cm]{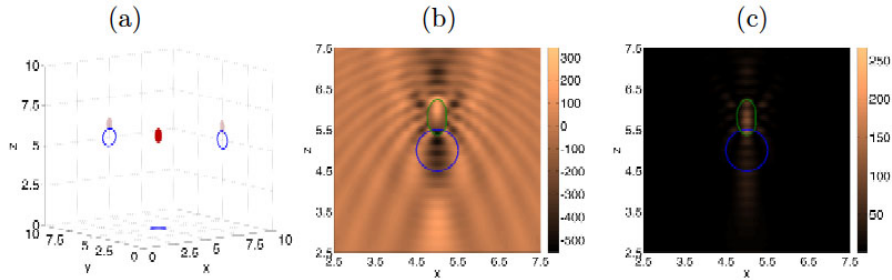}
\caption{Succesive approximations for the sphere of diameter 
$1 \, \mu$m.
(a) Initial guess $\Omega_0$. (b)-(c)
$y=5$ slices of  the topological derivative and  the topological
energy, respectively, computed with object $\Omega_0$ (green contour).
}
\label{fig13}
\end{figure}

We first revisit the pear-like object considered in Figure \ref{fig8}, illuminated with light of wavelength $660$ nm. Figure \ref{fig8} displays the initial topological derivatives and energies.
We use the topological derivative to define a first approximation through formulas (\ref{initialguesstd}) and (\ref{c0td})  with $\alpha_0=0.75$.
Figure \ref{fig12}(a) represents the first reconstruction $\Omega_0$ of the object. It is shifted towards the screen and contains points outside the true object. When computing the topological derivative for this first guess $\Omega_0$ of the true object, large positive values signal a top region to be removed and large negative values indicate new points to be added, see panel (d) in Figure \ref{fig12}.   We create the new object $\Omega_1$ using formulas (\ref{updatedguess2td})-(\ref{cC1}) with $\alpha_1=1/2$ and $\beta_1=1/80$. 
Panel (b) represents the new guess after the first  iteration. The reconstruction has moved towards the true location, correcting the shape. Computing the topological derivative for this new guess $\Omega_1$, we recover the elongated shape, see panel (e).

\begin{figure}[h!]
\centering
\includegraphics[width=12cm]{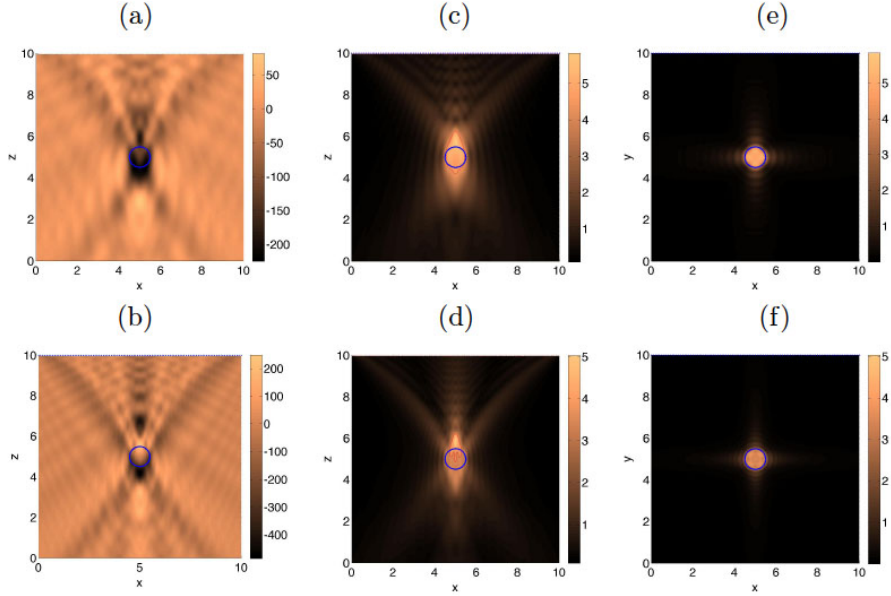}
\caption{ Topological derivatives and topological energies obtained
superimposing the fields corresponding to five incident waves
for the sphere of diameter $1 \, \mu$m,
one orthogonal to the detector screen and four symmetric ones
forming angles of amplitude $\pi/4$ with it.
(a)-(b)  Averaged topological derivatives at $y=5$ for the red and violet
lights, respectively.
(c)-(d)  Averaged square root of the topological energies at $y=5$ for the red and violet lights, respectively. (e)-(f) Same at planes $z=5$.
The blue contour of the true object is superimposed.}
\label{fig14}
\end{figure}

Next, we revisit the sphere of diameter $1$, located at $(5,5,5)$, illuminated with light  of wavelength  $660$ nm. The initial topological derivatives and  energies are shown in panels (h) and (k) of Figures \ref{fig5} and \ref{fig6}.
We use the topological energy to define a first approximation through formulas (\ref{initialguess}) and (\ref{c0}) with  $\alpha_0=0.75$. Figure \ref{fig13} displays the topological derivatives and topological energies computed after the first iteration.  Notice that only the bottom part of the initial object $\Omega_0$  remains in the next approximation $\Omega_1$, since the topological derivative is large  and positive  in the rest.  Moreover,  the topological derivative shows a prominent negative region below, covering the diameter of the true object. Other negative regions outside are an artifact associated to the spurious points included in $\Omega_0$. We can propose a new approximation $\Omega_1$ using either the peak of the topological energy (\ref{updatedguess2})
or the strategy (\ref{updatedguess2td}) for the topological derivative.
The computation of updated topological energies and derivatives in both cases does not yield a clear improvement. Instead, increasing the number of incident directions provides a much sharper reconstruction. In particular, we have analyzed the effect of superimposing the normalized TD and TE fields obtained adding four incident lights forming angles of $45$ degrees with the orthogonal one, in a symmetric way not to bias the information on the object shape. Some results for the ball of diameter $1 \, \mu$m are shown in Figure \ref{fig14}.
The TE fields seem to be easier to interpret than the TD fields, suggesting consistently the presence of an object about the true location, with accurate $xy$ sections, but slightly elongated in the $z$ direction.  Notice that the prediction is better for the red light, as shown by the  contours corresponding to a level $0.75 \,{\rm max}_{\mathbf{x}\in\mathcal{R}_{obs}} \, E_T({\bf x},{\mathbb R}^3)^{1/2}$. We have chosen to plot $E_T^{1/2}$ here
to better visualize the shape. The TE fields enhance the peaks and reduce
intermediate values. This effect is useful to distinguish different objects, but
may underestimate points around the peaks belonging to them. Single objects are sometimes more easily reconstructed using $E_T^{1/2}$, as it happens in this case.
The algorithm that reconstructs objects glueing together 2D sections along a suitable range $[z_1,z_2]$ described in Section \ref{sec:shapeidentification}, might produce comparable results, 
provided we are able to calibrate all the thresholds properly.

\section{Conclusions and future work}
\label{sec:discussion}

In microsystems such as cells, relevant structures span over a wide range of finer scales, down to a few nanometers. Non-invasive techniques which enable imaging biological samples over a range of spatial and temporal scales are of
great value. We have considered an imaging microsystem based on recording scattered light of single wavelengths on a screen.  This system displays a large range of effective wavenumbers depending on the geometry of the objects under study. Moreover, the possibility of employing several incident directions is restricted. Blending several wavelengths does not yield significative improvements either.  We have shown that  analytical formulas for the topological derivatives and energies of associated cost functionals provide good guesses of a variety of  three-dimensional configurations, 
based on the knowledge on the electric field at the detectors, the incident
light field and the background permittivity.

Using one incident wave, impinging orthogonally to the detector screen, we 
have discovered that topological methods allow us to locate  isolated objects with diameters down to $10$ nm  and trace their position with nanometer precision.
Lateral sections of the topological fields, orthogonal to the incidence direction, reproduce the form and size of the true objects for a variety of sizes and geometries. This would allow us to track shape changes with time. Asymmetric, non-convex and multiple contours are distinguishable.  Furthermore,  multiple scatterers distributed along axial and lateral directions are clearly identified.
On the other hand, axial information on single objects, in  the direction of the incident beam, must be interpreted taking into account the effective wavenumber, relative to the employed wavelength, and the object size.
Indeed, for  sizes above a micron, objects can be approximated connecting predictions of  two-dimensional sections along an axial range determined by 
two peaks in the topological fields, before and past the object. For larger wavelengths, or smaller sizes, the  two-dimensional  sections of interest lie along a peak of the topological field, usually with an offset towards the detector screen. This offset decreases as the size of the object diminishes. Besides,
 the elongation of the peak in the axial direction shrinks as the wavelength decreases, though the position may be more accurately described increasing the wavelength.

Assuming the permittivities of the objects known, we have tested 
different iterative procedures to refine initial reconstructions  keeping the 
same measured data, while improving the shape and position of 
the reconstructions. For small enough objects, topological derivative based iterations are suitable to approximate their size and shape in a  converging way.
In general, topological energies are usually more effective than topological derivatives to propose initial guesses because they are often easier to interpret in presence of oscillations, and the predicted position tends  to be  more accurate.  Moreover,  additional iterations based on topological energies may correct the offset. These refined guesses can also serve as starting point of alternative imaging algorithms that employ shape optimization or level set dynamics  to fit object contours, or gradient strategies to reconstruct
unknown permittivity profiles.

For specific choices of object sizes, depending on the distance to the screen, the area covered by detectors and the distance between them, the first reconstruction provided by the topological fields may be acceptable. Modifying these parameters to adapt them to the spatial scales of interest is another way of focusing the produced image.  Here,  we have analyzed the effect of incorporating a limited number of incident waves,  showing that this procedure largely improves the predictions of  position and size.  

Summarizing, this work demonstrates the potential of topological techniques in holographic microscopy.  However, our imaging setting assumes that we are able to measure the full wave field at the screen detectors. The data actually measured in holography are slightly different.
Both data types might be related as mentioned in the introduction, provided a way to estimate background fields is devised. Nevertheless, further work is required to be able to apply these reconstruction procedures in holographic microscopes. Proposing a deterministic (no or minimal human input needed) algorithm for turning topological fields into 3D models of objects would be of great value.   Our methods might be useful in other electromagnetic imaging problems, such as microwave imaging, where
electric fields are recorded and information from more sources is available.
\\

{\bf Acknowledgements.} A. Carpio thanks M.P. Brenner for hospitality while visiting Harvard University financed by a Caja Madrid Mobility grant.
T.G. Dimiduk and A. Carpio thank the Kavli Institute Seminars
at Harvard for the interdisciplinary communication environment that initiated this work. T.G. Dimiduk acknowledges the support of a National Science  Foundation (NSF) Graduate Research Fellowship.
This research was also supported by MINECO grants No. FIS2011-28838-C02-02 (AC), No. MTM2014-56948-C2-1-P  (AC, MLR), No. TRA2013-45808-R (MLR), No. MTM2013-43671-P (VS).
Part of the computations of this work were performed in EOLO,
the HPC of Climate Change of the Moncloa International Campus of Excellence, funded by MECD and MICINN.

\end{document}